\documentclass[10pt]{article}
\pdfoutput=1
\usepackage{graphicx}
\usepackage{jheppub}
\usepackage{amsfonts,amssymb}
\usepackage{amsmath}
\usepackage{epsfig}
\usepackage{bm}
\usepackage{hyperref}
\usepackage{color}
\usepackage{graphicx}	
\usepackage{epstopdf}
\newcommand{\f}{f}
\newcommand{\g}{g}
\def\p{\mathbf{p}}
\def\k{\mathbf{k}}

\usepackage{amsfonts,amssymb}
\usepackage{amsmath}

\makeatletter
\g@addto@macro\@floatboxreset\centering
\makeatother

\graphicspath{%
  {./},%
}

\begin{document}

\preprint{CERN-PH-TH-2015-304, MIT-CTP/4750}

\title{Weak and strong coupling equilibration in nonabelian gauge theories}

\author[a]{Liam Keegan}\affiliation[a]{Physics Department, Theory Unit, CERN, CH-1211 Gen\`eve 23, Switzerland}
\author[a,b]{Aleksi Kurkela}
\affiliation[b]{Faculty of Science and Technology, University of Stavanger
, 4036 Stavanger, Norway}
\author[c,d]{Paul Romatschke} 
\affiliation[c]{Department of Physics, 390 UCB, University of Colorado at Boulder, Boulder, CO, USA}
\affiliation[d]{Center for Theory of Quantum Matter, University of Colorado, Boulder, Colorado 80309, USA}
\author[e]{Wilke van der Schee} \affiliation[e]{Center for Theoretical Physics, MIT, Cambridge, MA 02139, USA}
\author[f]{and Yan Zhu} \affiliation[f]{University of Jyv\"askyla, Department of Physics, P.O. Box 35, FI-40014 University of Jyv\"askyl\"a, Finland}

\date{\today}

\abstract{
We present a direct comparison studying equilibration through kinetic theory at weak coupling and through holography at strong coupling in the same set-up. The set-up starts with a homogeneous thermal state, which then smoothly transitions through an out-of-equilibrium phase to an expanding system undergoing boost-invariant flow. This first apples-to-apples comparison of equilibration provides a benchmark for similar equilibration processes in heavy-ion collisions, where the equilibration mechanism is still under debate. We find that results at weak and strong coupling can be smoothly connected by simple, empirical power-laws for the viscosity, equilibration time and entropy production of the system.
}

\maketitle

\section{Introduction}

In the past decade, there has been a wealth of data on nuclear matter at extremely high temperatures from the experimental heavy-ion program at the Relativistic Heavy Ion Collider (RHIC) and the Large Hadron Collider (LHC) \cite{Adcox:2004mh,Arsene:2004fa,Adams:2005dq,Back:2004je,Aamodt:2010pa,Chatrchyan:2013nka,ATLAS:2012at}. To the surprise of many, hydrodynamic models are tremendously successful in describing and often predicting the experimental measurements \cite{Huovinen:2001cy,Romatschke:2007mq,Dusling:2007gi,Song:2007ux,Chojnacki:2007rq,Luzum:2008cw,Luzum:2009sb,Song:2010mg,Schenke:2010rr}.
However, one of the key requirements for the success of these hydrodynamic models is that the matter created after a relativistic ion collision equilibrates quickly, on a time-scale of $t_{\rm hydro}\sim 1-2$ fm/c  \cite{Luzum:2008cw,Heinz:2013wva}.
It has been a longstanding theoretical challenge to understand the pre-equilibrium physics that leads to this time scale. 

As a consequence, the search for a quantitative understanding of equilibration in gauge theories at high temperature has spawned a new subfield of physics. Two branches of this subfield have emerged, based on very different approaches. 
On the one hand, there is an effort to understand equilibration based on a weakly coupled framework. This branch was pioneered by the early parametric picture
of the so-called ``Bottom-up'' thermalization  \cite{Baier:2000sb}, and since then there has been a continuing effort to elevate the weak-coupling picture from parametric
estimates to a quantitative prescription by exploiting the scale separations provided by the 
weak coupling, admitting different effective theory descriptions \cite{Arnold:2002zm, Arnold:2005vb, Kurkela:2011ti,Kurkela:2011ub,Kurkela:2012hp,York:2014wja,Attems:2012js, Berges:2013eia,Berges:2013fga,Gelis:2013rba}. While the early parametric estimates 
were somewhat in tension with early thermalization \cite{Baier:2002bt}, the modern quantitative calculations are consistent with the fast thermalization \cite{Kurkela:2014tea,Kurkela:2015qoa}. 
The weak coupling approach can be rigorously set up for any gauge theory (such as ${\cal N}=4$ SYM and QCD \cite{Arnold:2002zm, Huot:2006ys}) whenever the coupling is small, but eventually will start to break down as the coupling is increased. 

On the other hand, equilibration at strong coupling has been studied using the gauge/gravity duality or holography \cite{Maldacena:1997re}, in which it is remarkably straightforward to study real time dynamics for certain gauge theories (such as ${\cal N}=4$ SYM, but not QCD). In holography the dynamics of the equilibration of the gauge theory is mapped onto the relaxation of a black hole in an anti-de-Sitter (AdS) space-time with one extra dimension  (see \cite{Chesler:2015lsa} for a recent review). Early studies near equilibrium suggested the black hole relaxes fast, with the characteristic time scale being $1/T$, with $T$ the temperature of the formed plasma \cite{Horowitz:1999jd}. In a non-linear setting the relaxation was pioneered by Chesler and Yaffe, who studied the relaxation of a gauge theory on a non-trivial curved background space-time \cite{Chesler:2008hg, Chesler:2009cy}. This was later extended to studies of a gauge theory in flat space-time, prepared with a wide variety of initial states \cite{Heller:2011ju,Heller:2012je,Heller:2012km,Heller:2013oxa}, which always led to hydrodynamics within $t<1.2/T$. Lastly, many studies have been performed in colliding settings, mimicking heavy-ion collisions more closely both in the longitudinal and transverse directions \cite{Kovchegov:2007pq,Grumiller:2008va,Balasubramanian:2010ce,Wu:2011yd,Heller:2012km,Casalderrey-Solana:2013sxa}, even allowing for direct comparison with experimental data \cite{vanderSchee:2013pia,vanderSchee:2015rta}.

For QCD at energy scales relevant for heavy-ion collisions the coupling constant is presumably not very small, nor very large. This makes it very interesting to compare the weakly and strongly coupled approaches, as they may bracket what happens in real heavy-ion collisions. There are several reasons why this comparison is not straightforward. In particular, the initial condition of the pre-equilibrium evolution at weak coupling is usually described in terms of classical fields or distribution functions whereas at strong coupling the initial condition has to be formulated in terms of fields in AdS space-time. In fact, it is not straightforward to characterise a far-from-equilibrium state in terms that are well defined and applicable in the both frameworks, which makes an apples-to-apples comparison of the evolution non-trivial. 

In this paper, we consider a setup which avoids the question of setting non-equilibrium initial conditions and allows a clean apples-to-apples comparison of the non-equilibrium evolution. We consider a system that is initially in thermal equilibrium but is subsequently pushed out of equilibrium by an external force. In practice we accomplish this by changing the metric rapidly with a pulse of curved space-time from a homogeneous Minkowski space to an expanding  space described by Milne coordinates. Finally, we compute the expectation value of the stress-energy tensor and the entropy density in both systems, such that we can follow how the system approaches a hydrodynamical description.

We find that at all values of the  't Hooft coupling $\lambda$ the system reaches hydrodynamical flow. For small couplings this is preceded by a period
of free-streaming type evolution which becomes shorter and shorter as the coupling is increased. As the coupling is increased the evolution starts to resemble the strongly coupled evolution. Indeed, for $\lambda=\infty$, the system is described by
hydrodynamics very quickly after the pulse has ended, but the departure from equilibrium does leave an imprint in non-equilibrium entropy production.

In section \ref{sec:one}, we explain our setup for driving gauge theories out of equilibrium, including a discussion on the evolution of the stress-energy tensor within hydrodynamics. For this setup, we describe state-of-the-art weak and strong coupling calculations in section \ref{sec:weak} and \ref{sec:strong}, and report our findings in section \ref{sec:results}. A summary and the conclusions that we draw from our findings can be found in section \ref{sec:conclusions}. 

\section{A simple set-up for studying gauge theory equilibration}
\label{sec:one}

In the following, we consider a gauge theory initially in global equilibrium at the temperature $T_i$. We then consider this gauge theory to be placed into a space-time with coordinates $x^a=(t,x,y,L)$ and line element
\begin{equation}
\label{eq:ds2}
ds^2=-dt^2+dx^2+dy^2+\g(t) dL^2
\end{equation}
with $\g(t)$ a function that smoothly transitions from $\g(t\rightarrow -\infty)=1$ to $\g(t\rightarrow \infty)\rightarrow t^2$ at late times $t$. This choice of metric tensor implies that for $t\rightarrow -\infty$, the gauge theory is in global equilibrium at rest within a flat space-time, as outlined above. By contrast, at late times, the gauge theory experiences stretching of (flat!) space-time in the longitudinal direction $z$. This late-time behavior corresponds to the familiar Bjorken flow \cite{Bjorken:1982qr}, since it is just a coordinate transformation of a gauge theory expanding in the longitudinal direction in Minkowski space. In between, the gauge theory experiences a dynamic, space-time pulse that is driving it (far) from its original equilibrium state. 

To be concrete, in the following we choose a one-parameter family of metric functions given by the choice
\begin{equation}
\label{eq:metricfunc}
\g(t)=\frac{1}{e^{\alpha(t T_i-1)}+1}+\frac{t^2 T_i^2}{e^{-\alpha (t T_i-1)}+1}\,,
\end{equation}
with $\alpha$ a free parameter controlling the rapidity of the change from early to late time behavior. It should be reiterated that for the choice of $g(t)$ in Eq.~(\ref{eq:metricfunc}), the space-time is flat up to exponentially small terms for all $t$ except the region $|t T_i-1|\propto \frac{1}{\alpha}$, where the space-time is curved.

It is possible to study the time-evolution of the stress-energy tensor components (such as the energy density) using hydrodynamic theory. The stress-energy tensor in hydrodynamic theory is given in terms of a gradient expansion. For a conformal theory, the most general stress tensor complete up to second order gradients in arbitrary $d$-dimensional space-times is given by \cite{ Baier:2007ix}
\begin{eqnarray}
\label{eq:BRSSS}
T^{\mu\nu}&=&\epsilon u^\mu u^\nu+P \Delta^{\mu\nu}+\pi^{\mu\nu}\,,\nonumber\\
\pi^{\mu\nu}&=&-\eta \sigma^{\mu\nu}+\eta \tau_\pi \left[^{\langle}D \sigma^{\mu\nu \rangle}+\frac{1}{d-1}\sigma^{\mu\nu}\left(\nabla \cdot u\right)\right]+\kappa
\left[R^{\langle \mu \nu\rangle}-(d-2)u_\alpha R^{\alpha <\mu\nu> \beta} u_\beta \right]\,\nonumber\\
&&+\lambda_1 \sigma^{<\mu}_{\quad \lambda}\sigma^{\nu> \lambda}+\lambda_2 \sigma^{<\mu}_{\quad \lambda}\Omega^{\nu> \lambda}+\lambda_3 \Omega^{<\mu}_{\quad \lambda} \Omega^{\nu> \lambda}\equiv \pi^{\mu\nu}_{BRSSS}\,,
\end{eqnarray}
where $\epsilon,P=\frac{\epsilon}{d-1}$ are the energy density and pressure for a conformal theory, $u^\mu$ is the fluid four-velocity (normalized to $u^\mu u_\mu=-1$), $\Delta^{\mu\nu}=g^{\mu\nu}+u^\mu u^\nu$ with $g^{\mu\nu}$ the metric tensor in the mostly plus sign convention and $R^{\mu\nu},R^{\alpha\mu\nu\beta}$ are the Ricci and Riemann tensors for this spacetime. Furthermore, the definition
$$
A^{\langle \mu\nu\rangle}\equiv \frac{1}{2}\Delta^{\mu\alpha}\Delta^{\nu\beta}\left(A_{\alpha \beta}+A_{\beta \alpha}\right)-\frac{1}{d-1} \Delta^{\mu\nu}\Delta^{\alpha \beta}A_{\alpha \beta}\equiv \langle A^{\mu\nu}\rangle\,
$$
has been used to define e.g. $\sigma^{\mu\nu}=2 \langle \nabla^\mu u^\nu\rangle$, where $\nabla^\mu$ is the covariant derivative for the metric $g^{\mu\nu}$. Moreover $D\equiv u^\mu \nabla_\mu$, and $\Omega^{\mu\nu}=\frac{1}{2}\Delta^{\mu\alpha}\Delta^{\nu\beta}\left(\nabla_\alpha u_\beta-\nabla_\beta u_\alpha\right)$. The coefficients $\eta,\tau_\pi,\lambda_1,\lambda_2,\lambda_3,\kappa$ are the first and second-order transport coefficients (material constants depending on the specific gauge theory and specific value of the coupling considered). In the limit where all of these transport coefficients are set to zero, one recovers ideal fluid dynamics, for which $\pi^{\mu\nu}=0$. In the limit where only the second-order transport coefficients $\tau_\pi,\kappa,\lambda_1,\lambda_2,\lambda_3$ are set to zero one recovers
\begin{equation}
\label{eq:NS}
\pi^{\mu\nu}=-\eta \sigma^{\mu\nu}\equiv \pi^{\mu\nu}_{NS}\,,
\end{equation}
which is the constitutive relation of the relativistic Navier-Stokes equation. We will refer to the equations (\ref{eq:NS}) and the expression for $T^{\mu\nu}$ as Navier-Stokes (NS) theory, while the full set of equations (\ref{eq:BRSSS}) will be referred to as BRSSS in the following.

It turns out that as they stand, both the NS equations (\ref{eq:NS}) and the BRSSS equations (\ref{eq:BRSSS}) would be acausal, as can be easily seen by working out the group velocity from the dispersion relations \cite{Romatschke:2009im}. It has proven very useful for practical applications to consider the following \emph{resummed} version of the BRSSS constitutive equations, which will be referred to as rBRSSS in the following:
\begin{eqnarray}
\label{eq:rBRSSS}
\pi^{\mu\nu}&=&-\eta \sigma^{\mu\nu}-\tau_\pi \left[^{\langle}D \pi^{\mu\nu \rangle}+\frac{d}{d-1}\pi^{\mu\nu}\left(\nabla \cdot u\right)\right]+\kappa
\left[R^{\langle \mu \nu\rangle}-(d-2)u_\alpha R^{\alpha <\mu\nu> \beta} u_\beta \right]\,\nonumber\\
&&+\lambda_1 \sigma^{<\mu}_{\quad \lambda}\sigma^{\nu> \lambda}+\lambda_2 \sigma^{<\mu}_{\quad \lambda}\Omega^{\nu> \lambda}+\lambda_3 \Omega^{<\mu}_{\quad \lambda} \Omega^{\nu> \lambda}\,\equiv \pi^{\mu\nu}_{rBRSSS}\,.
\end{eqnarray}
The rBRSSS equations are generally causal, stable and with various choices of transport coefficients have been forming the backbone of modern hydrodynamic modeling of experimentally probed heavy-ion collisions (see e.g. \cite{Luzum:2008cw,Schenke:2010rr}). Note that resummed versions of hydrodynamic equations have appeared in the literature before, for instance the well-known Israel-Stewart equations \cite{Israel:1979wp} that is often employed in heavy-ion literature. In the present context we note that because we consider perturbations away from flat space-time, the Israel-Stewart equations would not be sufficient to treat the system evolution up to second-order gradients because the space-time described by Eq.~(\ref{eq:metricfunc}) is not flat, and terms proportional to $\kappa$ in Eq.~(\ref{eq:rBRSSS}) would be missing, distorting the system evolution.

Let us now consider the case of $d=4$ in the following. For the line element (\ref{eq:ds2}), with the initial condition of global equilibrium at temperature $T_i$, one finds that the fluid dynamic solution maintains the initial condition of vanishing spatial flow velocity, so that $u^\mu=({1,\bf 0})$. This implies $\epsilon=T^{L}_L$, and effective transverse and longitudinal pressures of $P_\perp\equiv T^{x}_x=T^y_y=P \left(1+2 H\right)$, $P_L\equiv T^L_L=P \left(1-4 H\right)$, respectively, with $H\equiv \frac{-\pi^L_L}{\epsilon+P}$. For later convenience, it is useful to define the pressure anisotropy as
\begin{equation}
\label{eq:plpt}
\frac{P_L}{P_\perp}=\frac{1-4 H(t)}{1+2 H(t)}\,.
\end{equation}

The covariant conservation of the stress-energy tensor $u_\mu \nabla_\nu T^{\mu\nu}=0$ leads to
\begin{equation}
\label{eq:master}
\partial_t \ln s = -\frac{\g^\prime(t)}{2 \g(t)}\left(1-H(t)\right)\,,
\end{equation}
where $s=\frac{\epsilon+P}{T}=\frac{4 \epsilon}{3 T}$ is the \emph{equilibrium} entropy density and $T(t)$ is the temperature. It is convenient to express dynamic quantities with respect to their initial (global equilibrium) values,
\begin{equation}
\frac{s(t)}{s_i}\equiv \frac{T^3(t)}{T_i^3}\,,\quad
\frac{\epsilon(t)}{\epsilon_i}\equiv \frac{T^4(t)}{T_i^4}\,,\quad {\rm etc.}
\end{equation}

For ideal hydrodynamics ($\pi^{\mu\nu}=0$ and thus $H=0$), the conservation of energy can be solved analytically in closed form for an arbitrary metric function $\g(t)$:
\begin{equation}
\label{eq:IHsol}
\frac{s_{\rm ideal}(t)}{s_i}=\g^{-1/2}(t)\,,\quad
\frac{\epsilon_{\rm ideal}(t)}{\epsilon_i}=\g^{-2/3}(t)\,,\quad
\left. \frac{P_L}{P_\perp}\right|_{\rm ideal}=1\,.
\end{equation}
Plots of the ideal hydrodynamic solution for the metric function (\ref{eq:metricfunc}) will be shown in section \ref{sec:results}. For further reference, it is useful to define the concept of the total equilibrium entropy $S_{eq}$ in the system, defined as
\begin{equation}
\label{eq:Seqdef}
\frac{S_{eq}}{S_{eq,i}}\equiv\frac{\int d^3x \sqrt{-{\rm det}g(t)} s(t)}{\int d^3x s_i}=\g^{1/2}(t) \frac{s(t)}{s_i}\,,
\end{equation}
which for ideal hydrodynamics trivially becomes $\frac{S_{eq,ideal}(t)}{S_{eq,i}}=1$. This just reflects the fact that no entropy is created in ideal (inviscid) hydrodynamics. Note that the equilibrium entropy thus defined will only correspond to the total system entropy if the system is close to equilibrium. Otherwise, non-equilibrium (viscous) corrections to the equilibrium entropy cannot be neglected (see e.g. the discussion in Ref.~\cite{Romatschke:2009kr}).

Including viscous corrections, one finds for Navier-Stokes hydrodynamics
\begin{equation}
H_{NS}(t)=\frac{2}{3}\frac{\eta}{s}\frac{\g^{\prime}(t)}{\g(t) T(t)}\,,
\end{equation}
which has to be solved together with (\ref{eq:master}). 
Using the fact that $g(t)$ really is a function of $t\, T_i$ only, the equations of motion for Navier-Stokes become
\begin{equation}
\label{eq:NSdyn1}
\partial_t \ln \frac{s}{s_i}=-\frac{\g^\prime}{2 \g}\left(1-\frac{2 \eta }{3 s}\frac{\g^\prime}{\g  T_i} \left(\frac{s}{s_i}\right)^{-1/3}\right)\,.
\end{equation}

Going beyond Navier-Stokes, for the case at hand the rBRSSS equations reduce to (\ref{eq:master}) plus the dynamic constitutive equation
\begin{equation}
\label{eq:rBRSSSeq}
\tau_\pi \partial_t H(t)=-\frac{4}{3}\frac{\eta}{s}\frac{\g^\prime}{2 \g T}-H
-\frac{4}{3} \tau_\pi T H \left(\partial_t \ln s+\frac{\g^\prime}{2 \g}\right)
-\frac{\lambda_1}{\eta^2} \frac{H^2 s T}{2}+\frac{\kappa}{sT}
\frac{\g^{\prime 2}-2 \g \g^{\prime \prime}}{\g^2}\,.
\end{equation}
In the following, the common parametrizations
\begin{equation}
\tau_\pi=\frac{C_\tau \eta}{s T}\,,\quad
\lambda_1=\frac{C_\lambda \eta^2}{s T}\,,\quad
\kappa=\frac{C_\kappa s}{T}\,,
\end{equation}
will be used.

\subsection{Hydrodynamic late time limit}

In the late time limit, when $\g(t)\simeq t^2$, the Navier-Stokes equations (\ref{eq:NSdyn1}) can be solved as a gradient expansion around the ideal hydrodynamic solution. Assuming a constant value of $\frac{\eta}{s}$, one finds
\begin{equation}
\label{eq:NStot}
\left.\frac{s(t)}{s_i}\right|_{ NS, t T_i\gg 1}=\frac{\chi}{t T_i}\left(1-\frac{2 \eta}{s} \frac{1}{(t T_i)^{2/3} \chi^{1/3}}\right)\,,
\quad
\left.\frac{P_L}{P_\perp}\right|_{ NS,t T_i\gg 1}=1-\frac{8 \eta}{s}\frac{1}{(t T_i)^{2/3} \chi^{1/3}}\,.
\end{equation}
Unlike solutions to the full evolution equations (\ref{eq:NSdyn1}), which assume that Navier-Stokes hydrodynamics is accurate through the entire time-evolution, the late-time solution (\ref{eq:NStot}) is a universal solution to the system evolution for all initial conditions, and as such includes an unknown constant $\chi$. Comparing the total equilibrium entropy resulting from (\ref{eq:NStot}) to that from ideal hydrodynamics, one can interpret 
\begin{equation}
\chi = \frac{S_{eq}(t\rightarrow \infty)}{S_{eq,i}}
\end{equation}
 as the total amount of  entropy produced. 
 
Similarly, one can also go beyond Navier-Stokes and solve the BRSSS equations of motion using a gradient expansion. One finds
\begin{equation}
\label{eq:BRSSStot}
\left.\frac{s(t)}{s_i}\right|_{BRSSS,t T_i\gg 1}=\frac{\chi}{t T_i}\left(1-\frac{2 \eta}{s}\frac{1}{(t T_i)^{2/3} \chi^{1/3}}+\frac{2\eta^2(2+C_\lambda-C_\tau)}{3 s^2 (t T_i)^{4/3}\chi^{2/3}} \right)\,,
\end{equation}
where it should be noted that the constant $C_\kappa$ does not enter the result because the space-time is flat for $\g(t)\rightarrow t^2$ at late times. In a similar fashion one finds
\begin{equation}
\label{eq:BRSSSplpt}
\left.\frac{P_L}{P_T}\right|_{BRSSS,t T_i\gg 1}=1-\frac{8 \eta}{s}\frac{1}{(t T_i)^{2/3} \chi^{1/3}}+\frac{16 \eta^2(3+C_\lambda-C_\tau)}{3 s^2 (t T_i)^{4/3} \chi^{2/3}}\,.
\end{equation}

\section{Gauge theory dynamics from a weak coupling approach}
\label{sec:weak}

In the non-interacting limit $\lambda \rightarrow 0$, the system is described by non-interacting free-streaming particles whose
evolution is given by the collisionless transport equation for the on-shell particle distribution $\f$
\begin{equation}
p^\mu \partial_\mu \f - \Gamma^{i}_{\alpha \beta} p^{\alpha}p^\beta \frac{\partial \f}{\partial p^i}=0, 
\end{equation}
where the summation of $i$ goes over the spatial coordinates $x,y$ and $L$. The transport equation with the initial condition and metric we have chosen, becomes
\begin{equation}
\label{eq:KEfs}
\partial_t \f - \frac{\g'}{ \g} p^L \frac{\partial \f}{\partial p^L} = 0\,.
\end{equation}

Eq.~(\ref{eq:KEfs}) can be solved analytically for arbitrary metric function $g(t)$ using the method of characteristics, and one obtains $f=f(p^x,p^y,p^L \sqrt{g(t)})$ (cf. Ref~\cite{Romatschke:2015dha}). With an initially thermal system with temperature $T_i$, a full solution to the free-streaming evolution of bosons is thus given by 
\begin{equation}
\label{eq:FSsol}
f=\frac{1}{\exp{\left[\frac{\sqrt{(p^{x})^2+(p^{y})^2+g^{2}(t)\, (p^{L})^2}}{T_i}\right]}-1}\,.
\end{equation}
The solution (\ref{eq:FSsol}) can be brought into the form $f=\sum_{n=1}^\infty
\exp\left[-\frac{n \sqrt{{\bf p}^2\left(1+\xi(t) \left(\frac{p^L}{|{\bf p}|}\right)^2\right)}}{T_i}\right]$, where 
\begin{equation}
\xi(t)=g(t)-1
\end{equation} 
is the anisotropy parameter defined in Ref.~\cite{Romatschke:2003ms}. This identification allows direct connection to anisotropic plasma physics literature, and in particular leads to the expressions for the energy density and pressure anisotropy \cite{Strickland:2014pga}: 
\begin{equation}
\label{eq:FSsols}
\frac{\epsilon_{FS}(t)}{\epsilon_i}=\frac{1}{2}\left(\frac{1}{1+\xi(t)}+\frac{{\rm arctan}\sqrt{\xi(t)}}{\sqrt{\xi(t)}}\right)\,,\qquad
\left.\frac{P_L(t)}{P_\perp(t)}\right|_{FS}=2 \frac{(1+\xi(t))\frac{\epsilon_{FS}(t)}{\epsilon_i}-1}{1-(1-\xi^2(t))\frac{\epsilon_{FS}(t)}{\epsilon_i}}\,,
\end{equation}
for a system of free-streaming (FS) particles experiencing arbitrary metric perturbations of the form (\ref{eq:ds2}). In the late-time limit, these lead to the following expressions for equilibrium entropy and pressure anisotropy
\begin{equation}
\frac{s_{FS,t T_i\gg 1}(t)}{s_i}=\left(\frac{\pi}{4 t T_i}\right)^{4/3}\,,\qquad
\left.\frac{P_L(t)}{P_\perp(t)}\right|_{FS,t T_i\gg 1}=\frac{2}{(t T_i)^2}\,,
\end{equation}
which upon comparison with Eqns.~(\ref{eq:NStot}) imply that the system never reaches equilibrium. Thus, free-streaming (non-interacting) evolution is the opposite extreme to ideal hydrodynamic evolution, and one expects these two extreme cases to bound the system evolution for any interaction strength $\lambda \in (0,\infty)$.

For numerical purposes, it is useful to work with rescaled longitudinal momentum $p^z=\sqrt{g(t)} p^L$ such that $f(t,p^x,p^y,p^L)\rightarrow \hat f(t,p^x,p^y,p^z)=\hat f(t,p^x,p^y,\sqrt{g} p^L)$. There is a Jacobian 
$$
\frac{\partial\left. f\right|_{p^L}}{\partial t}= \frac{ \partial \left.\hat f\right|_{p^L}}{\partial t}=\frac{\partial \left.\hat f\right|_{p^z}}{\partial t}+\frac{\partial f}{\partial p^z}\frac{p^L \partial \sqrt{g(t)} }{\partial t}
$$ 
associated with this transformation (cf. Ref.~\cite{Romatschke:2011qp}) which gives rise to the rescaled equation
\begin{equation}
\label{eq:KEfs2}
\partial_t \hat \f - \frac{\g'}{2\g} p^z \frac{\partial \hat \f}{\partial p^z} = 0\,.
\end{equation}
(It is easy to check this new form by using the free-streaming solution (\ref{eq:FSsol}), or simply putting $\hat f\propto (p^x)^2+(p^y)^2+g (p^z)^2$). Since it is of no importance, we denote $\hat f\rightarrow f$ in the following even though it is a distribution evaluated at constant $p^z$, not constant $p^L$.

At small but finite coupling, most of the modes can still be described by a transport equation,
which due to collisions among particles will now be of the form (\ref{eq:KEfs}), but with a non-vanishing right-hand side. We will restrict our weak coupling simulations to SU($N$) gauge theory. Then assuming that the non-equilibrium distributions are not spin polarized, the transport equation for the spin (and color) independent distribution function of gauge bosons can be written in a form where the right hand side contains two effective collision terms that contribute to leading order in $\lambda$ \cite{Arnold:2002zm}, one describing elastic ($2\leftrightarrow 2$), and the other inelastic ($1\leftrightarrow 2$) particle interactions:
\begin{equation}
\label{eq:fullKT}
\partial_t \f - \frac{\g'}{2 \g} p^z \frac{\partial \f}{\partial p^z} = -\mathcal{C}^{2\leftrightarrow 2}-\mathcal{C}_a^{1\leftrightarrow 2}.
\end{equation}
The collision operators read
\begin{align}
\label{2to2}
 \mathcal{C}_{2\leftrightarrow 2}[\f](\p)&= \frac{1}{4|\p| \nu}\int_{\k \p' \k'}|\mathcal{M}(\p,\k;\p',\k')|^2 (2\pi)^4 \delta^{(4)}(P+K-P'-K') \nonumber \\
 &\times \big\{ \f(\p)\f(\k)[1+ \f(\p')][1+ \f(\k')]-\f(\p')\f(\k')[1+ \f(\p)][1+ \f(\k)] \Big\}
\end{align}
and
\begin{align}
\label{1to2}
\mathcal{C}^{1\leftrightarrow 2}[\f](\p) &= \frac{(2\pi)^3}{2|\p|^2 \nu}\int_0^{\infty} dp' dk' \delta(|\p|-p' -q')\gamma(\p;p'\hat{\p},k'\hat{\p})\nonumber \\
&\hspace{1cm}\times
\big\{ \f(\p)[1+ \f(p' \hat{\p})][1+ \f(k' \hat{\p})] - \f(p' \hat{\p})\f(k' \hat{\p})[1+ \f(\p)]\Big\}\nonumber\\
&+ \frac{(2\pi)^3}{|\p|^2 \nu}\int_0^{\infty} dp' dk \delta(|\p|+k -p')\gamma(p' \hat{\p};\p,k \hat{\p}) \nonumber\\
&\hspace{1cm}\times
\big\{ \f(\p)\f(k \hat{\p})[1+ \f(p' \hat{\p})] -\f(p' \hat{\p})[1+ \f(\p)][1+ \f(k\hat{\p})]\Big\},
\end{align}
where $\nu= 2 d_A$ is the number of degrees of freedom.
Here $|\mathcal{M}|^2$ and $\gamma$ are effective matrix elements for elastic scattering and collinear splitting, respectively.
Both of them have non-trivial structure arising from the soft and collinear divergences of the underlying processes which are 
regulated dynamically by in-medium physics. The soft $t$ and $u$ channel divergences present in vacuum are regulated by
physics of screening at scale $m^2 = \lambda \int \frac{d^3 p}{(2\pi)^3} \f(\p)/|\p|$, whereas the collinear singularity in splitting term
gets regulated through physics of Landau-Pomeranchuk-Migdal suppression \cite{Migdal:1956tc,Landau:1953um,Landau:1953gr}. In practice these effects are included by performing Hard-Loop and ladder resummations in the required kinematic regions. A detailed discussion of effective matrix elements can be found in \cite{Arnold:2002zm,York:2014wja,Kurkela:2015qoa}. In the current work we will restrict ourselves to  isotropic screening approximation introduced in \cite{Kurkela:2015qoa}, such that our description is strictly accurate to leading order in $\lambda$ only when the system is isotropic $P_L/P_T\simeq 1$. For anisotropic systems the prescription gives a description that is accurate to 
leading logarithm order.  For systems with $P_L/P_T -1 \gtrsim \lambda^{1/3}$ certain plasma unstable modes may start to give a parametrically leading order contribution \cite{Kurkela:2011ti}, which are neglected in the isotropic screening approximation. Even though the effect of the plasma instabilities is parametrically
a leading order contribution, classical Yang-Mills simulations suggest that their numerical effect is still negligible \cite{Romatschke:2006nk,Berges:2013eia}. Within this approximation the effective matrix element $|\mathcal{M}|^2$ reads
\begin{align}
|\mathcal{M}|^2 = 8 \lambda^2 \nu \left( \frac{9}{4} + \frac{(s-t)^2}{\bar{u}^2}+ \frac{(u-s)^2}{\bar{t}^2}+ \frac{(t-u)^2}{s^2}\right),
\end{align}
where the $\bar{t}$ and $\bar{u}$ are regulated Mandestam variables. For $t \rightarrow 0$, the matrix element is proportional to $\propto 1/(q^2)^2$ where $q = | {\bf p'}-{\bf p}|$ is the momentum transfer in the elastic collision. In the isotropic screening approximation we regulate the soft $t$-channel (and similarly for the $u$-channel) divergence by replacing 
\begin{equation}
q^2 \bar{t} \rightarrow \left( q^2 +2 \xi_0^2 m^2 \right) t
\end{equation}
in the denominator of the divergent terms. The coefficient $\xi_0 = e^{5/6}/\sqrt{8}$ is chosen such that the collision term 
reproduces the drag and momentum diffusion properties of the soft scattering at leading order \cite{York:2014wja}. 

The effective splitting rate $\gamma$ is given by
\begin{align}
\gamma(p {\bf \hat p}; p' {\bf \hat p},k' {\bf \hat p}) = \frac{p^4 + p'^4+k'^4}{p^3 p'^3 k'^3 }\frac{\nu \lambda}{4 (2\pi)^3}\int \frac{d^2 h}{(2\pi)^2} 2 {\bf h} \cdot {\rm Re} {\bf F},\nonumber
\end{align} 
where the equation for ${\bf F}$ accounts for splitting due to multiple scatterings with transverse momentum exchange ${\bf} q$, and momentum non-collinearity ${\bf h} = {\bf p}\times {\bf k}$
\begin{align}
&2 {\bf h} =  i \delta E({\bf h} ){\bf F}({\bf h})+\frac{\lambda T_*}{2}\int \frac{d^2  q_\perp}{(2\pi)^2} \mathcal{A}({\bf q}_\perp)  \Big[ 3{\bf F({\bf h})}-{\bf F}({\bf h}-p {\bf q}_\perp)-{\bf F}({\bf h}-k {\bf q}_\perp)-{\bf F}({{\bf h}+p' {\bf q}_\perp})\Big]\nonumber.
\end{align}
with
$
T_* = \frac{\lambda}{2 m^2}\int \frac{d^3 p}{(2\pi)^3}f_\p(1+f_\p) ,
$
and $\delta E = m^2/p'+m^2/k'-m^2/p + {\bf h}^2/2p k' p'$. In the isotropic screening approximation
$$
\mathcal{A}({\bf q}_\perp) = \left( \frac{1}{{\bf q}_\perp^2} - \frac{1}{{\bf q}_\perp^2 + 2m^2}\right).
$$
This integral equation is most conveniently solved by the powerful numerical method introduced in 
\cite{Ghiglieri:2014kma}.

Note that both $|M|^2$ and $\gamma$ are proportional to $\nu$ so that dependence on the number of colors $N_c$ enters only though the definition of the 't Hooft coupling $\lambda =g^2 N_c$. Therefore the evolution of distribution functions are independent of $N_c$ to the order considered here. 

In order to numerically solve the transport equation, we impose an azimuthal symmetry to the distribution function so that
it becomes a function of the absolute value of momentum $p=|\p|$ and the polar angle $x = \cos(\hat \p \cdot \hat {\bf z})$, that is $f(\p) = f(x,p)$. 
We then discretize the continuous distribution $f(x ,p)$ by introducing
a 2-dimensional grid $\{x_i,p_j\}$ in $x$ and $p$ labeled by the indices $i$ and $j$. In our implementation we choose to keep track of the number densities $n_{ij}$ near the grid points and define a discretized variable 
\begin{equation}
n_{ij} = \int \frac{d^3 p}{(2\pi)^3} f(x,p) w_{ij}(x,p)\,,
\end{equation}
where we have also defined the 2-dimensional \emph{wedge function} $w_{ij}$ which is centered around the grid point  labelled by indices $i$ and $j$
\begin{align}
w_{ij}(x,p) &=w_i(x) w_j(p) \\
w_a(z) &= \frac{z_{a+1}-z}{z_{a+1}-z_a}\theta(z-z_a)\theta(z_{a+1}-z)+\frac{z-z_{a-1}}{z_{a}-z_{a-1}}\theta(z_a-z)\theta(z-z_{a-1}), \quad a=i,j.
\end{align}
In terms of discretized number densities the transport equation reads
\begin{align}
\frac{d n_{ij}}{dt} + C_{ij}^{\rm exp} = - C_{ij}^{2\leftrightarrow 2} - C_{ij}^{1\leftrightarrow 2} \label{dnijdt},
\end{align}
where $ C_{ij}^{2\leftrightarrow 2}$ and $ C_{ij}^{1\leftrightarrow 2}$ are the discretized collision operators and $C_{ij}^{\rm exp}$ is 
the discretized version of the derivative arising from the non-trivial metric. 

We have a freedom in choosing the discretization $C_{ij}^{\rm exp}$.  
In continuum, the evolution of local energy density and particle number density due to the non-trivial metric 
is exactly related to the components of the energy momentum tensor by partial integration identities. In the absence of
interactions the time evolution of the energy and number densities are
\begin{align}
\frac{2g}{g'} \frac{d \epsilon}{d t} & = \nu  \int \frac{d^3 p}{(2\pi)^3} p \partial_t f = - \nu  \int  \frac{d^3 p}{(2\pi)^3} p p^z \partial_{p^z} f = - \nu  \int  \frac{d^3 p}{(2\pi)^3} (p + \frac{(p^z)^2}{p}) = - (\epsilon + P_L); \label{dedt}\\
\frac{2g}{g'} \frac{d n}{d t} &= \nu  \int \frac{d^3 p}{(2\pi)^3} \partial_t f =  -\nu  \int  \frac{d^3 p}{(2\pi)^3} p^z \partial_{p^z} f = -n \label{dndt}.
\end{align}
Both collisional terms conserve energy density exactly and therefore Eq.~(\ref{dedt}) holds also in the presence of interactions. Inelastic processes on the other hand change particle number and Eq.~(\ref{dndt}) receives a contribution from $\mathcal{C}_{1\leftrightarrow 2}$ in the interacting case. In our numerical implementation we choose to discretize $C_{ij}^{\rm exp}$ so that it exactly reproduces Eqs.~(\ref{dedt},\ref{dndt}). In terms of discretized quantities the different integral moments read
\begin{align}
n = \nu \sum_{ij} n_{ij}, \quad \epsilon = \nu \sum_{ij} p_j n_{ij} \quad \text{and}  \quad P_L = \nu \sum_{ij} \frac{x_i^2 p_j^2}{p_j} n_{ij},
\end{align}
and we can write the first order discretized derivative operator as
\begin{align}
\frac{2 g}{g' }C^{\rm exp}_{ij}
\equiv   \phantom{+} & n_{ij}  x_i^2 \left[ \frac{p_{j} }{p_j - p_{j-1}}\right] -   n_{i,j+1} x_i^2 \left[ \frac{p_{j+1} }{p_{j+1}-p_{j}}\right] \\
 + &n_{ij} \left[ \frac{(x_{i-1} - x_{i-1}^3)}{x_i - x_{i-1}} \right] - n_{i+1,j}\left[ \frac{(x_{i} - x_{i}^3)}{x_{i+1} - x_{i}} \right] - n_{ij}.
\end{align}

The distribution function itself is needed to compute the Bose enhancement factors in Eqs.(\ref{2to2}) and (\ref{1to2}). In terms of the 
discretized quantity, the distribution function is approximated by $\tilde f$
which is defined by 
\begin{align}
\frac{4\pi p^2}{(2\pi)^3} \tilde{f}(x,p) &\equiv  \sum_{ij} \frac{n_{ij} w_{ij}(x,p)}{v_{ij}},\\
v_{ij} &= \int \frac{d^3 p}{(2\pi)^3} w_{ij}(x,p).
\end{align}
In terms of discretized variables, the collision terms then read
\begin{align}
C_{ij}^{1\leftrightarrow 2} =&\frac{v_i}{\nu} \int_0^{\infty} dp \int_0^{p/2} dp'  \int_0^{\infty }dk' \delta(p-p'-k') [4\pi\gamma(p;p',k')
] \\ &
\times\{ \tilde f(x_i,p)[1+\tilde f(x_i,p')][1+ \tilde f(x_i,k')]-\tilde f(x_i,p') \tilde f(x_i,k')[1+\tilde f(x_i,p)]\} \nonumber 
  \left[ w_{ij}(x_i,p)-w_{ij}(x_i,p')-w_{ij} (x_i,k')\right]
\end{align}
with $v_i = \int_0^{1} d x w_i(x)$ and 
\begin{align}
C_{ij}^{2\leftrightarrow 2}=&\frac{1}{8 \nu}\int d\Gamma_{PS} |\mathcal{M}|^2 \{  \tilde f(x_p,p) \tilde f(x_k,k) [1+  \tilde f(x_{p'},p')][1+ \tilde f(x_{k'},k')]-\tilde f(x_p,p) \tilde f(x_k,k) [1+\tilde f(x_{p'},p')][1+\tilde f(x_{k'},k')] \}\nonumber \\
\times&\left[ w_{ij}(x_p,p)+ w_{ij}(x_k,k)- w_{ij}(x_{p'},p')- w_{ij}(x_{k'},k')\right].
\end{align}
Because our interpolation with the wedge functions exactly reproduces linear functions, $\sum_{ij} w_{ij}(x,p) p = p$, 
the discretized collision kernels exactly conserve
energy. The integral over the phase space of $2\leftrightarrow 2$ scatterings reads
\begin{align}
\int d\Gamma_{\rm PS} \equiv \frac{1}{2^{11}\pi^7}&\int_0^\infty dq \int_{-q}^{q}d\omega \int_{(q-\omega)/2}^{\infty}
dp \int_{(q+\omega)/2}^{\infty}dk \nonumber \int_{-1}^{1} d x_q \int_{0}^{2\pi}d \phi_{pq}d \phi_{kq}, \label{gamma_PS}
 \end{align}

with $p' = p+w $ and $k' = k-w$.  In terms of these coordinates, the angles of incoming and outgoing momenta needed for the arguments of the occupation numbers are given by
\begin{equation}
x_{\{p\}}= -\sin\theta_{\{p\}q}\cos\phi_{\{p\}q} \sqrt{1-x_q^2}+\cos\theta_{\{p\}q}x_q\;,
\end{equation}
where $\{ p\} = p,k,p',k'$ with $\cos \phi_{p'q} = \cos \phi_{pq}$,  $\cos \phi_{k'q} =  \cos \phi_{kq}$. The cosines appearing in the previous formula are given by
\begin{align}
\cos\theta_{pq} = \frac{\omega}{q} + \frac{t}{2 pq}, & \quad 
\cos\theta_{kq} = \frac{\omega}{q} - \frac{t}{2 kq},\\
\cos\theta_{p'q} = \frac{\omega}{q} - \frac{t}{2 p' q}, & \quad
\cos\theta_{k'q} = \frac{\omega}{q} + \frac{t}{2 k' q},
\end{align}
and $t \equiv \omega^2 - q^2$.  The effective matrix element $|\mathcal{M}|^2$ depends also on Mandelstam $s$ and $u$ which in terms of the integration variables read
\begin{align}
s &= 
\frac{-t}{2 q^2 } \left\{ 
(p+p')(k+k')+q^2 -  \cos(\phi_{kq}-\phi_{pq})\sqrt{(4 p p' + t)(4 k k'+t) }\right\}
\\
u &= -t -s.
\end{align}

In our numerical implementation, we start with the initial condition $f(p) = 1/(e^{p/T_i} -1)$ at some early time $t_i < 0$. We then 
use a simple time stepping algorithm to determine the distribution function at later times by iterating 
\begin{equation}
n_{ij}(t+\Delta t ) = n_{ij}( t ) - \Delta t \left(C_{ij}^{\rm exp}+C_{ij}^{2\leftrightarrow 2}+C_{ij}^{1\leftrightarrow 2}\right) \label{timestep}.
\end{equation}
In order to evaluate the collision terms, we first measure at each timestep the values of the thermal mass $m$ and the 
effective temperature $T_*$
\begin{equation}
m^2 = \lambda \sum_{ij} \frac{n_{ij}}{p_j}, \quad \quad T_*= \frac{\lambda}{2 m^2} \sum_{ij} n_{ij}(1 + \tilde f(x_i,p_j))
\end{equation}
and then estimate the collision kernels $C_{ij}^{2\leftrightarrow 2}$ and $C_{ij}^{1\leftrightarrow 2}$ by Monte Carlo sampling the phase space integrals. We have found that it is essential to use importance sampling that reflects the important regions in the phase space. In particular we sample the integral over $q$ with the weight $dq/q^2$ that accounts for the soft divergence in $|\mathcal{M}|^2$, whereas the soft divergence appearing in $C^{1\leftrightarrow 2}_{ij}$ is ameliorated by  sampling the $p'$ integral with the weight $dp'/p'$.

\section{Gauge theory dynamics from a strong coupling approach}
\label{sec:strong}

At strong coupling we use holography to study the process described in section \ref{sec:one}. This is done in the simplest version of holography, which allows to describe processes in $\mathcal{N}=4$ super-Yang-Mills theory (SYM) through dynamics of Einstein gravity in 5 dimensional anti-de-Sitter (AdS) spacetime. We are hence led to solving Einstein's equations in AdS, whereby the non-trivial geometry (Eqn. (\ref{eq:ds2})) corresponds to non-trivial boundary conditions on AdS. The starting condition with a thermal state corresponds to a black brane geometry in AdS.

In this paper we will use the characteristic formulation of Einstein's equations, first introduced in \cite{Bondi1960Gravitational, Sachs,Bondi}, and later extended to AdS in \cite{Chesler:2008hg, Chesler:2009cy}. The essential ingredient of this formulation is the metric ansatz, which is written in null coordinates and has the spatial determinant ($S$) factored out:
\begin{equation}
ds^{2}=2dtdr-Adt^{2}+S^{2}e^{B}dx^{2}+S^{2}e^{B}dy^{2}+S^{2}e^{-2B}dL^{2}\,,\label{eq:metricEF}
\end{equation}
where $A$, $S$ and $B$ are functions of time $t$ and the AdS radial
coordinate $r$. This coordinate choice for the metric makes the Einstein equations particularly simple \cite{Chesler:2008hg}:\begin{subequations}
\begin{eqnarray}
0 & = & S''+{\textstyle \frac{1}{2}}B'^{2}\, S\,,\label{Ct}\\
0 & = & S\,(\dot{S})'+2S'\,\dot{S}-2S^{2}\,,\label{eq:Seq}\\
0 & = & S\,(\dot{B})'+{\textstyle \frac{3}{2}}\big(S'\dot{B}+B'\,\dot{S}\big)\,,\label{Beq}\\
0 & = & A''+3B'\dot{B}-12S'\,\dot{S}/S^{2}+4\,,\label{Aeq}\\
0 & = & \ddot{S}+{\textstyle \frac{1}{2}}\big(\dot{B}^{2}\, S-A'\,\dot{S}\big)\,,\label{Cr}
\end{eqnarray}
\label{Eeqns} \end{subequations} where 
\begin{equation}
h'\equiv\partial_{r}h\quad\mathrm{and}\quad\dot{h}\equiv\partial_{t}h+{\textstyle \frac{1}{2}}A\,\partial_{r}h\label{eq.defdirderivs}
\end{equation}
denote derivatives along the ingoing and outgoing radial
null geodesics, respectively.  As can be seen, given some initial metric, specified in our case by the field $B(r,\,t=t_0)$, we can integrate the first four equations successively, after which we have obtained $\partial_t B(r,\,t=t_0)$, which allows to step forward in time. 

These integrations require boundary conditions, which we obtain from the near-boundary analysis of the Einstein equations. For this we need to specify the metric of the boundary of AdS, which according to the AdS/CFT dictionary should equal the metric of the CFT. This means that $S^{2}e^{B}=r^2$ and $S^{2}e^{-2 B}=g(t) r^2$ to leading order in $r$, which results in the following asymptotic forms:
\begin{eqnarray} A(r,t) & = & r^2+\frac{\left(g'\right)^2-2 g g''}{6 g^2}+\frac{288 a_4 g^4+ \log (\text{r})\left(8 g''' g^2 g'-4 g^2 \left(g''\right)^2+7 \left(g'\right)^4 )-12 g \left(g'\right)^2 g'' \right)}{288 g^4 r^2}+O\left(r^{-3}\right),\\
 B(r,t) & = & -\frac{\log (g)}{3}-\frac{g'}{3 g r}+\frac{3 \left(g'\right)^2-2 g g''}{24 g^2 r^2}+\frac{3 g''' g^2-\left(g'\right)^3-4 g g' g''}{108 g^3 r^3}+\\
& & \frac{576 b_4 g^4+\log(\text{r})\left( 34 g''' g^2 g' -12 g'''' g^3 +28 g^2 \left(g''\right)^2 +35 \left(g'\right)^4 -84 g \left(g'\right)^2 g'' \right)}{576 g^4
   r^4}+O\left(r^{-5}\right), \\
 S(r,t) & = & \sqrt[6]{g} r+\frac{g'}{6 g^{5/6}}-\frac{\left(g'\right)^2}{36 g^{11/6} r}+\frac{4 \left(g'\right)^3-3 g g' g''}{324 g^{17/6} r^2}+\frac{72 g''' g^2 g'-36 g^2 \left(g''\right)^2-65
   \left(g'\right)^4-12 g \left(g'\right)^2 g''}{31104 g^{23/6} r^3}+O\left(r^{-4}\right)\,,\nonumber
\end{eqnarray}
where $g(t)$ is given by eqn. (\ref{eq:metricfunc}), and $a_4(t)$ and $b_4(t)$ depend on the complete bulk dynamics and cannot be fixed by a near-boundary analysis. In our actual computation we computed the $\log(r)$ terms to a high order ($O(r^{-8})$ for $B(r,t)$) to deal with these analytically and thereby stabilize our numerics. Note that we fixed the remaining gauge freedom of $r \rightarrow r + \xi(t)$  in these expansions by the leading term in $S$. We later used this gauge freedom $\xi(t)$ to fix our coordinates such that the apparent horizon starts and remains at $r = 1$.

Formally, the stress tensor of our SYM theory is equal to the variation of the AdS action with respect to the boundary metric. This, however, is divergent, and just as in the SYM theory this has to be renormalized by adding appropriate counter-terms to the action. This procedure is known as holographic renormalization, and is carried out in \cite{deHaro:2000xn,Bianchi:2001de}. In our case this leads to:
\begin{eqnarray}
 \epsilon  & = & \frac{N_c^2}{2 \pi ^2} \left( -\frac{3 a_4}{4}-\frac{\left(\left(g'\right)^2-2 g g''\right)^2}{768 g^4}+\mu  \epsilon_{\text{SD}} \right)\,, \\
 P_{\perp} & = & \frac{N_c^2}{2 \pi ^2} \left( -\frac{a_4}{4}+b_4+\frac{608 g''' g^2 g'+4 g^2 \left(107 \left(g''\right)^2-48 g'''' g\right)+515 \left(g'\right)^4-1388 g
   \left(g'\right)^2 g''}{6912 g^4}+\mu  P_{\text{SD}, \perp}\right)\,,\nonumber \\
 P_L & = & \frac{N_c^2}{2 \pi ^2} \left( -\frac{a_4}{4}-2 b_4+\frac{-1168 g''' g^2 g'+384 g'''' g^3-844 g^2 \left(g''\right)^2-979 \left(g'\right)^4+2668 g \left(g'\right)^2 g''}{6912
   g^4} +\mu  P_{\text{SD}, L} \right)\,,\nonumber
\end{eqnarray}
where the terms proportional to $\mu$ depend on the renormalization scheme, i.e. due to the presence of the conformal anomaly when the boundary metric is curved it is possible to add finite counterterms to the action, with a coefficient $\mu$ that needs to be fixed by a choice of scheme \cite{deHaro:2000xn,Bianchi:2001de,Papadimitriou:2004ap}. For our metric these are given by:
\begin{eqnarray}
 \epsilon _{\text{SD}} & = & \frac{8 g''' g^2 g'-4 g^2 \left(g''\right)^2+7 \left(g'\right)^4-12 g \left(g'\right)^2 g''}{384 g^4}, \\
 P_{\perp} & = & \frac{-20 g''' g^2 g'+8 g'''' g^3-20 g^2 \left(g''\right)^2-21 \left(g'\right)^4+52 g \left(g'\right)^2 g''}{384 g^4}, \\
 P_{\text{SD},L} & = & \frac{48 g''' g^2 g'+4 g^2 \left(9 \left(g''\right)^2-4 g'''' g\right)+49 \left(g'\right)^4-116 g \left(g'\right)^2 g''}{384 g^3}\,. 
\end{eqnarray}
Note, however, that these terms are absent in the case where the boundary metric is flat (when $g(t)=1$ or when $g(t)=t^2$), so that these terms are only important around our pulse, near $t T_{ini}\simeq 1$. Also, all contributions are fourth order in derivatives, which is why we did not have to take the scheme dependence into account when considering 2nd order hydrodynamics in section \ref{sec:one}.

We are now able to numerically solve the Einstein equations (see also \cite{Chesler:2013lia,vanderSchee:2014qwa} for a more detailed discussion), where we started our evolution at $t\, T_{ini}=-10$, with $A=r^2 - (\pi T)^4/r^2$, $S=r$ and $B=0$. As already alluded to, the near-boundary behavior of the metric functions is handled analytically, where we subtracted many of the logarithmic terms for increased stability. The spatial discretization is then done using spectral elements\cite{boyd2001chebyshev}, using 6 domains with 15 grid points, and for time stepping we used an explicit Adams-Bashforth scheme \footnote{The \emph{Mathematica} code to evolve an evolution as described is available upon request at \href{mailto:wilke@mit.edu}{wilke@mit.edu}; alternatively, simpler versions can be found at \href{https://sites.google.com/site/wilkevanderschee}{sites.google.com/site/wilkevanderschee}}.

\begin{figure}[t]
\includegraphics[width=0.99\linewidth]{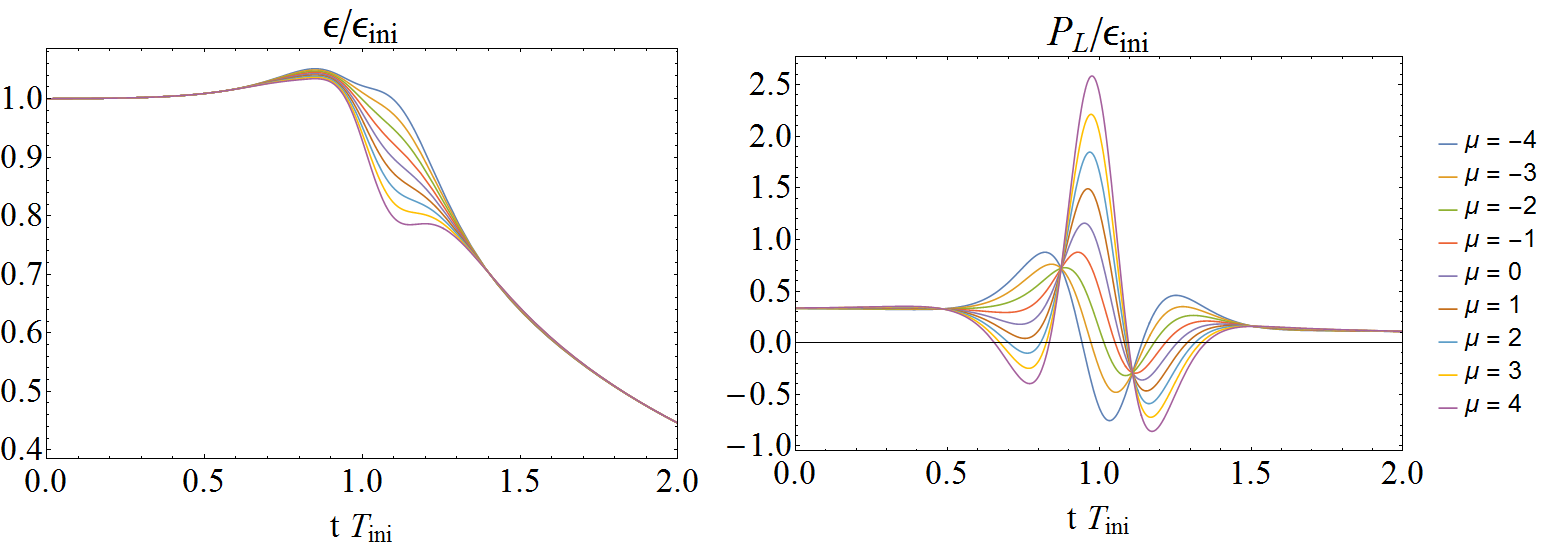}
\caption{\label{fig:scheme} For the energy density and transverse pressure we illustrate for $\alpha=8$ in eqn. (\ref{eq:metricfunc}) the renormalization scheme dependence present for the expectation value of stress-energy tensors of QFTs living in an even-dimensional curved spacetime. Clearly, the pressure oscillates wildly during the period where the boundary metric is curved (as also noted in \cite{Chesler:2008hg}), but these oscillations are mostly due to scheme dependence. We therefore focus on the times after the pulse ($t\,T_i \gtrsim 1.4$), where this ambiguity does not arise, and present results for the choice of $\mu=-2$.}
\end{figure}

Finally, having obtained the full AdS metric, we can extract the normalizable modes of the metric ($a_4(t)$ and $b_4(t)$) to obtain the expectation value of the SYM stress tensor. As an illustration of the scheme dependence presented above we plot $\epsilon$ and $P_L$ in figure \ref{fig:scheme}, for several values of $\mu$ and for $\alpha=8$. The choice of scheme is clearly important around $t\, T_{ini}$, but is unimportant in the expanding regime after the pulse (with $f(t)=t^2$). For this particular boundary metric a reasonable choice is $\mu=-2$, which leads to the mildest oscillations possible (though for different $\alpha$ different $\mu$ would be preferred in that sense). For the results to be presented in the next section we hence used $\mu = -2$.

Lastly, as in the hydrodynamic and weak coupling approaches, we also kept track of a measure of entropy. Here we used the area density ($S^3$) of the apparent horizon, which location is given by $\dot{S}(t,\,r_{AH})=0$. While this measure depends on the time slicing of the AdS metric, it can be determined locally in time (as opposed to the event horizon, which depends on the full future spacetime). Also, this time slicing ambiguity in the definition of the entropy can be compared with similar ambiguities in a field theory far-from-equilibrium \cite{Booth:2011qy}.

\section{Results}
\label{sec:results}

\begin{figure}[t]
\includegraphics[width=0.65\linewidth]{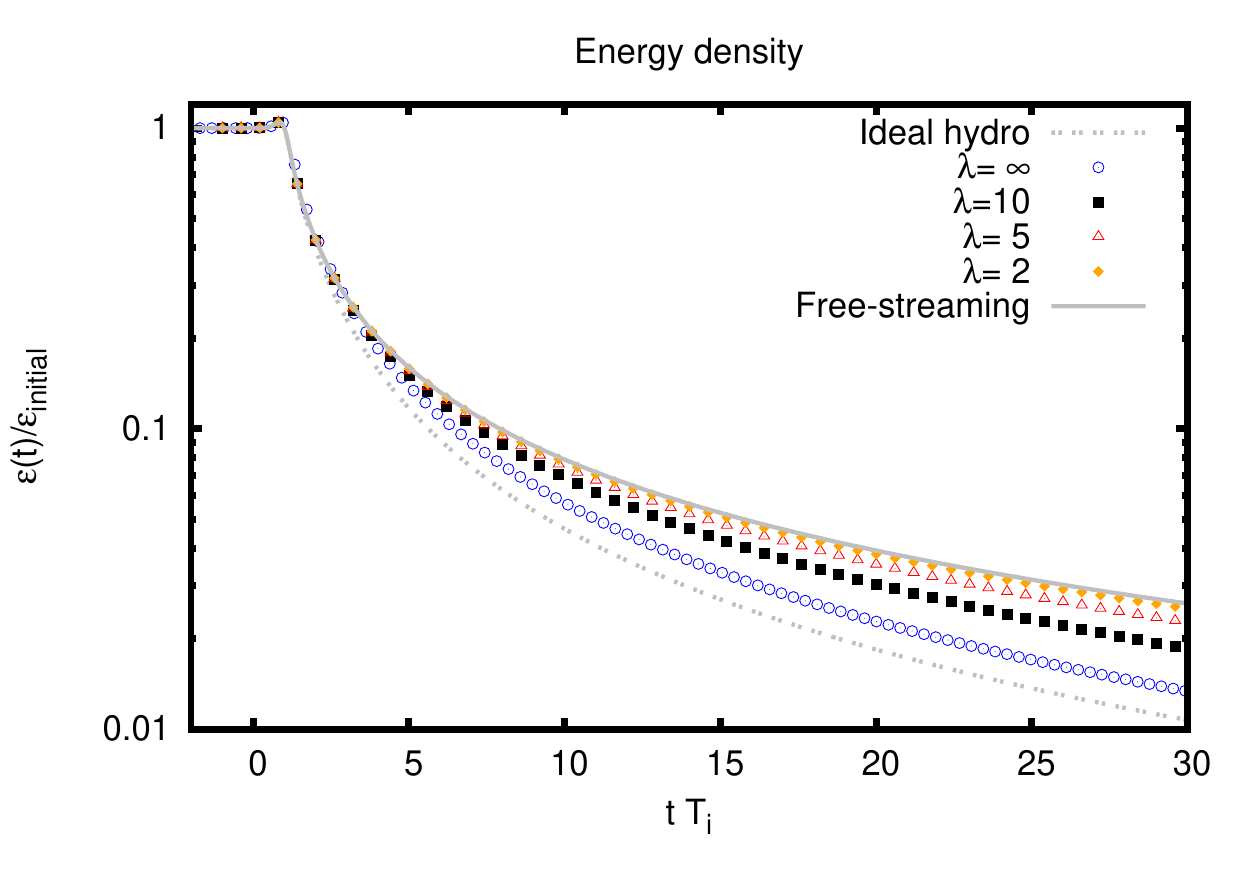}
\caption{\label{fig:one} Time evolution of the energy density from kinetic theory ($\lambda=2,5,10$) and the gauge/gravity duality ($\lambda=\infty$). For reference, the analytic results for non-interacting particles ($\lambda=0$, ``free-streaming'') and ideal hydrodynamics (``ideal hydro'') are also plotted.}
\end{figure}

\begin{figure}[t]
\includegraphics[width=0.5\linewidth]{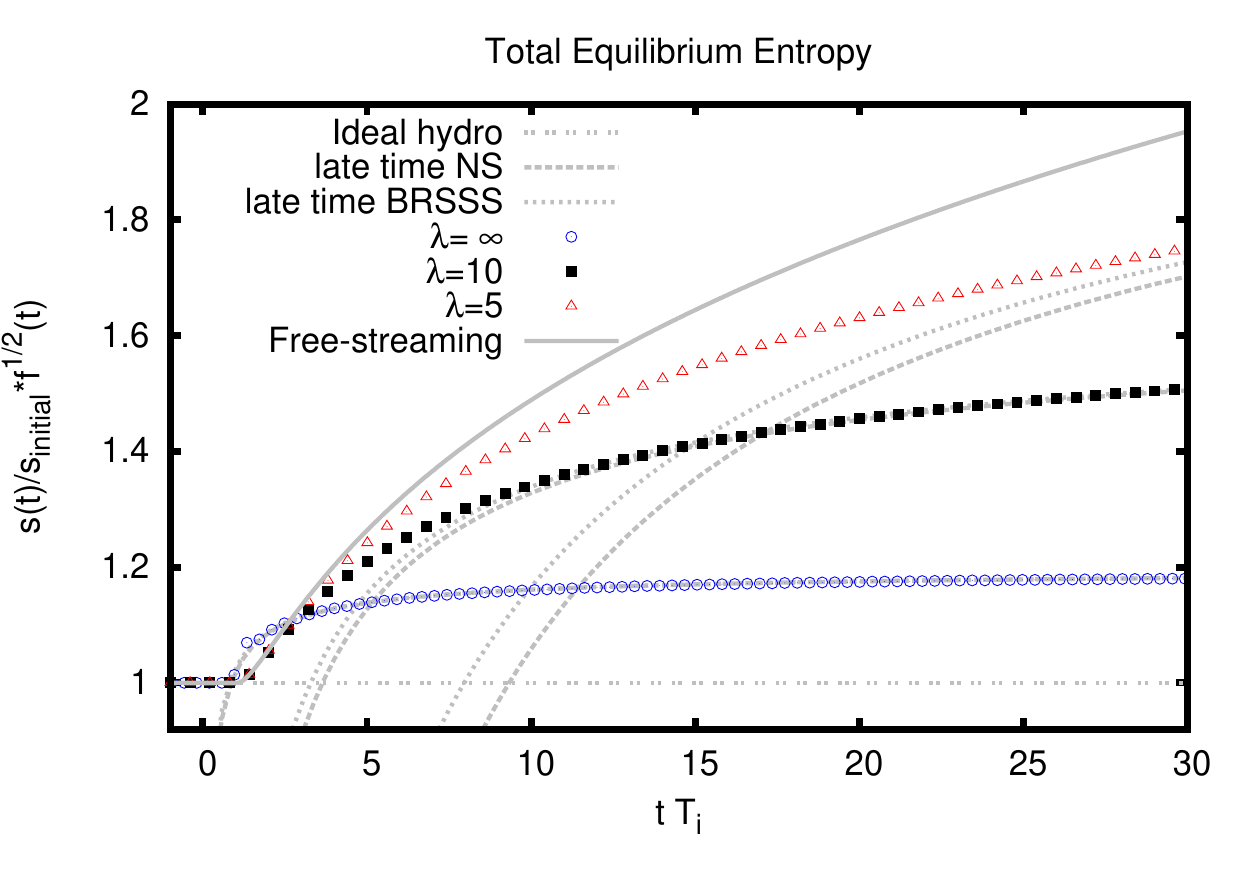}\hfill
\includegraphics[width=0.5\linewidth]{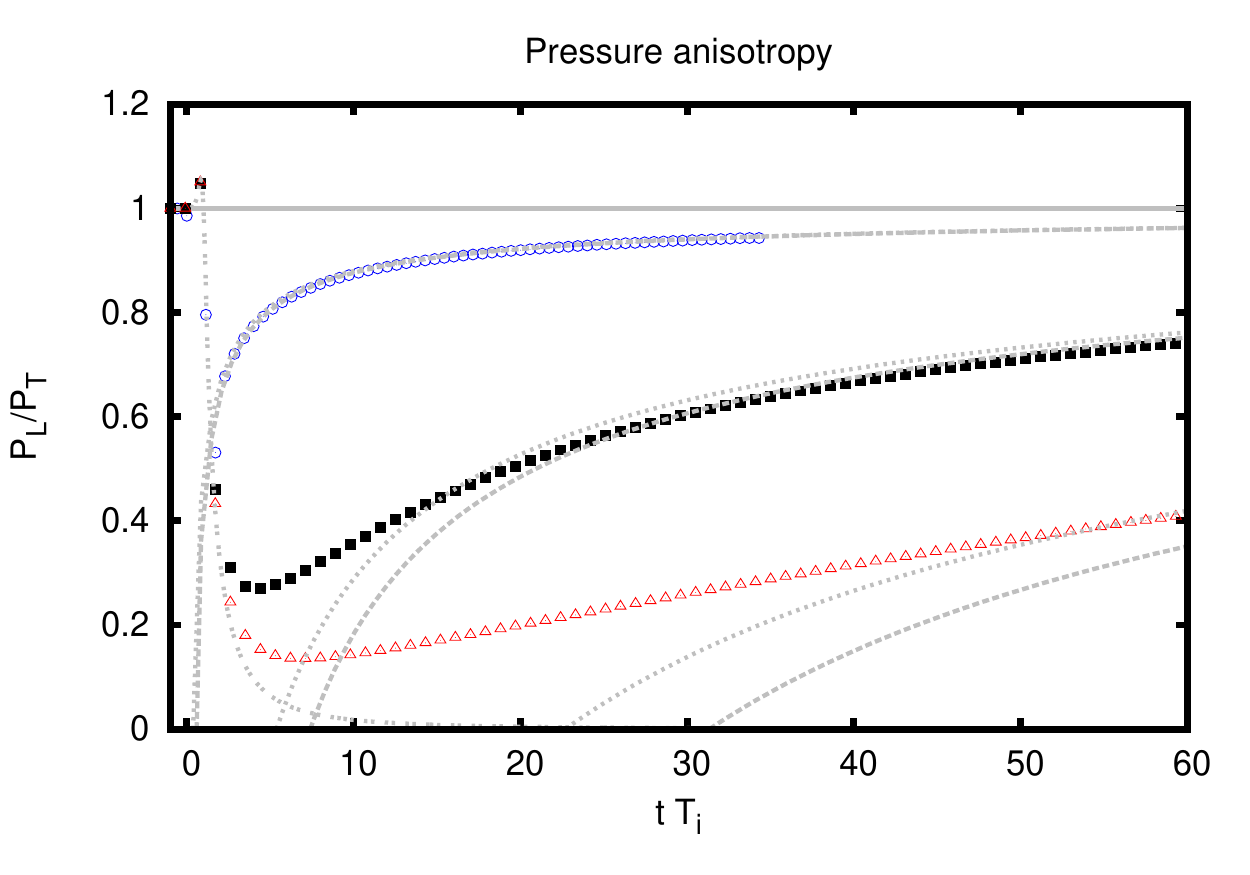}
\caption{\label{fig:two} Time evolution of the total system equilibrium entropy (left) and pressure anisotropy (right). Shown are results from kinetic theory ($\lambda=5,10$) and gauge/gravity duality ($\lambda=\infty$). For reference, the analytic results for non-interacting particles ($\lambda=0$, ``free-streaming''), ideal hydrodynamics (``ideal hydro'') as well as the late-time gradient expansion to first-order (NS) and second order (BRSSS) hydrodynamics with transport coefficients from Tab.\ref{tab:one} are also shown. }
\end{figure}

As described in the introduction, the system is prepared in a thermal initial state at time $t_0<0$ and then subjected to the boundary metric pulse of Eq.~(\ref{eq:ds2}). For weak coupling $\lambda={\cal O}(1)$, the evolution of the system is solved using the kinetic theory methods described in section \ref{sec:weak}, while for strong coupling $\lambda\rightarrow \infty$, gauge gravity methods described in section \ref{sec:strong} were used. In Fig.~\ref{fig:one}, the time evolution of the energy density is shown for the case of a pulse profile given by Eq.~(\ref{eq:metricfunc}) with $\alpha=8$. As expected, the energy density of the system drops at late times consistent with the expansion of the system. For decreasing values of the coupling $\lambda\sim 1$, the kinetic theory simulation approaches the analytic free-streaming result given in Eq.~(\ref{eq:FSsols}). For strong coupling $\lambda \rightarrow \infty$, the result is somewhat closer to the analytic ideal hydrodynamics result Eq.~(\ref{eq:IHsol}) than the kinetic theory result for intermediate coupling $\lambda=10$, but does not coincide with the ideal hydrodynamic result since viscous corrections do not vanish even for $\lambda=\infty$.

This difference to ideal hydrodynamics is highlighted when plotting the total equilibrium entropy and pressure anisotropy, defined in Eqns.~(\ref{eq:Seqdef}, \ref{eq:plpt}), as done in Fig.~\ref{fig:two}. In this figure, results for $\lambda=5,10,\infty$ are plotted along with the ideal hydrodynamics and free-streaming results. 
For the equilibrium entropy one finds that with the exception of the non-interacting case (free-streaming), all curves tend to a constant value for $t\rightarrow \infty$, which quantifies the amount of entropy production during the evolution process. Determining the asymptotic value of entropy corresponds to fixing the parameter $\chi$ in Eqns.~(\ref{eq:NStot}, \ref{eq:BRSSStot}). 
For the pressure anisotropy we clearly see that the systems equilibrate towards isotropy, which allows us to define an isotropization time $t_{\rm iso}$ as the last time when $P_L/P_T= 0.8$.

Both panels in Fig.~\ref{fig:two} also include the curves which follow from late time hydrodynamics, both for first order (``NS'') and second-order (``BRSSS'') hydrodynamics, as given by Eqns.~(\ref{eq:NStot}) and (\ref{eq:BRSSSplpt}) respectively, whereby we use the transport coefficients from Tab.~\ref{tab:one} \footnote{The values of $\eta/s$ in Table \ref{tab:one} have been extracted from the late behaviour of stress-energy tensor in our current setup. The values 
agree with the original calculation of \cite{Arnold:2003zc} within 10\%. The two calculations differ from each other in the way the soft divergence is regulated. Both of these calculations are accurate to leading order but differ at subleading orders in $\lambda$, and therefore correspond to different possible definitions of leading order. The small discrepancy between two results can be understood as an estimate of the systematic theory uncertainty introduced in the kinetic theory at finite $\lambda$. }. While the evolution for $\lambda=5 \text{ and } 10$ shows that at sufficiently early times the hydrodynamic and kinetic results are clearly different, it is somewhat surprising to see that for $\lambda=\infty$, there seems to be almost perfect matching after the metric pulse has passed at $t T_i\simeq 1$. This seems to indicate that for $\lambda=\infty$, the system never actually leaves thermal equilibrium for the type of perturbation studied here (cf. the discussion in Ref.~\cite{Bhattacharyya:2009uu}).

\begin{figure}[t]
\includegraphics[width=0.65\linewidth]{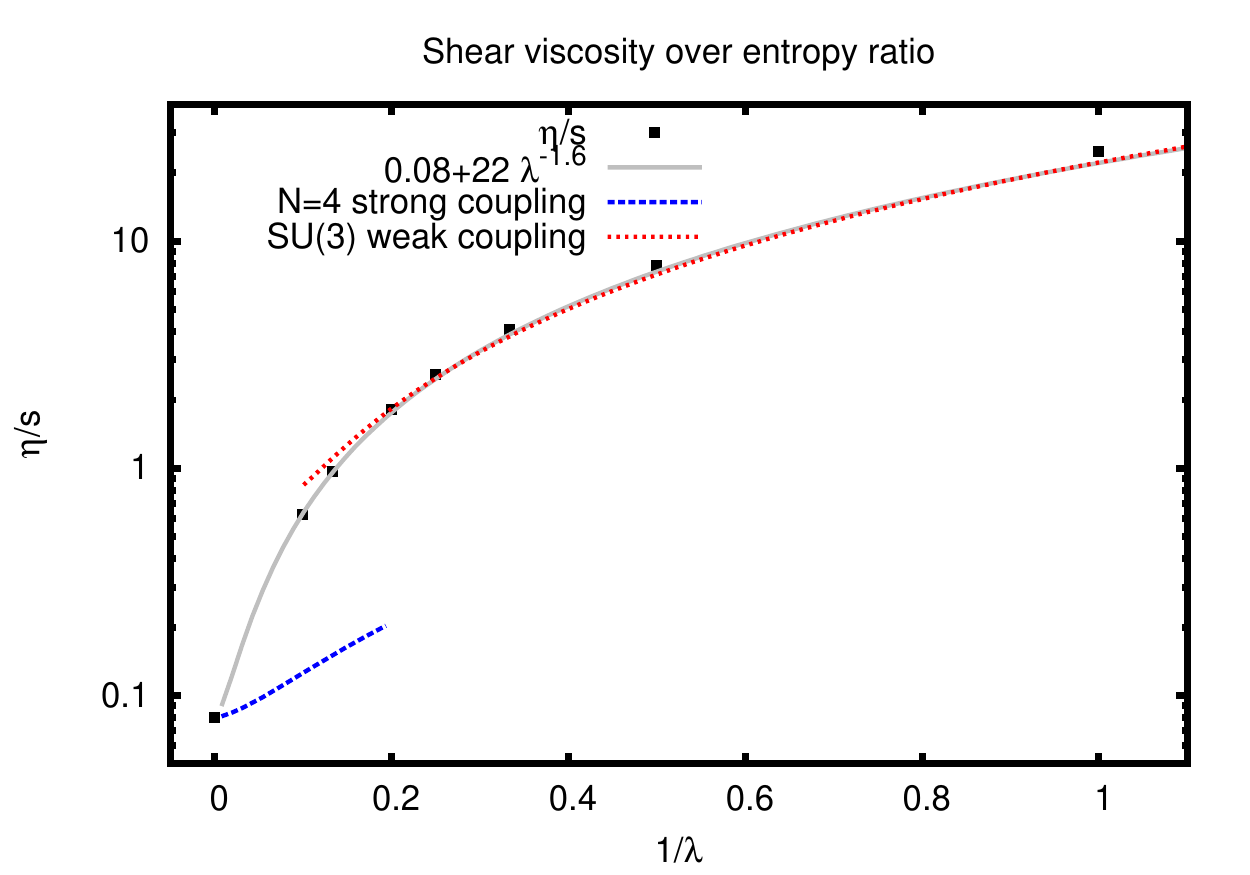}
\caption{\label{fig:eta} Shear viscosity over entropy density for weakly coupled SU(3) \cite{Arnold:2003zc}, for strongly coupled ${\cal N}=4$ SYM \cite{Policastro:2001yc,Buchel:2008sh}, compared to the values in Table \ref{tab:one} and the empirical interpolation formula (\ref{eq:etaform}). See text for details.}
\end{figure}

\begin{table}
\begin{tabular}{|c|c|c|c|c|c|c|c|c|}
\hline
 \text{} & \text{$\lambda $ = 1.0} & \text{$\lambda $ = 2.0} & \text{$\lambda $ = 3.0} & \text{$\lambda $ = 4.0} & \text{$\lambda $ = 5.0} & \text{$\lambda $ = 7.5} & \text{$\lambda $ = 10.0} & \text{$\lambda $ = $\infty $} \\
\hline
 \text{$\eta /s$} & 24.7 & 7.84 & 4.09 & 2.59 & 1.81 & 0.966 & 0.624 & 0.0796 \\
\hline
 $C_{\tau }$ & 5.4 & 5.3 & 5.3 & 5.2 & 5.2 & 5.2 & 5.1 & 2.6 \\
\hline
$C_\lambda$ &     4.5 & 4.3     & 4.3    & 4.2   & 4.2 & 4.1 & 4.1      & 2\\
\hline
$t_{\rm iso} (\eta/s)^{-4/3}$ & 176 & 174 & 176 & 178 & 178 & 180 & 181 & 142 \\
\hline
$t_{\rm NS} T_i $ & 6740 & 387 & 204 & 126.5 & 83& 37&21& 2.1\\
\hline
$t_{\text{BRSSS}}T_i $ & 7596 &418&120&73&48&23 & 13.5 & 1.7\\
\hline
$ \chi  $& 6.29 & 4.00 & 3.13 & 2.66 & 2.36 & 1.97 & 1.71 & 1.20 \\\hline
\end{tabular}
\caption{\label{tab:one} Summary of transport parameters and derived quantities for SU($N$) gauge theory for various values of $\lambda$ as well as for $\mathcal{N}=4$ SYM for $\lambda = \infty$. The $\eta/s$ values are extracted from the late time behaviour of the energy momentum tensor, while the second order parameters are taken from Refs.~\cite{Kovtun:2004de,York:2008rr,Bhattacharyya:2008jc}.  $\chi = S_{eq}(t\rightarrow \infty)/S_{eq,i}$ is the total (original plus viscously produced) entropy and $t_{iso}$ is the isotropization time. $t_{NS,BRSSS}$ refer to equilibration times from first and second-order hydrodynamics, respectively (see text for details).  }
\end{table}

In order to quantitatively study how well the system is described by hydrodynamics we define the first  ($t_{NS}$), and second order ($t_{BRSSS}$) hydrodynamization times as the time when the first or second order late time hydrodynamic result agrees with the $P_L/P_T$ within some fiducial range. As the deviations are not monotonic, we also demand that the hydrodynamical expression is within the fiducial range at all later times, whereby we take this range to be 5\%. We report the values of all times and transport coefficients mentioned above in Table \ref{tab:one}. We now first explore the scaling with $\lambda$ of the various quantities extracted, after which we plot a (rescaled) version of Fig. \ref{fig:two} in Fig.  \ref{fig5}, in order to highlight the observed trends.

In Fig. \ref{fig:eta} the $\eta/s$ values are plotted as a function of the coupling $\lambda$. We find that the analytic leading-log formula of weak coupling SU(3) from Ref.~\cite{Arnold:2003zc},
$$
\left.\frac{\eta}{s}\right|_{\rm SU(3),\lambda \ll 1}=\frac{34.784}{\lambda^2 \log\left[4.789/\sqrt{\lambda}\right]}
$$ 
accurately captures the kinetic theory result up to $\lambda\lesssim 5$. The kinetic theory results from Table \ref{tab:one} nicely connect to the ${\cal N}=4$ SYM result for $\lambda\rightarrow \infty$ using the empirical interpolation formula
\begin{equation}
\label{eq:etaform}
\frac{\eta}{s}\simeq 0.08+22 \lambda^{-1.6}\,.
\end{equation}
However, the result for ${\cal N}=4$ SYM including
strong coupling corrections from Ref.~\cite{Buchel:2008sh},
$$
\left.\frac{\eta}{s}\right|_{\rm {\cal N}=4, \lambda\gg 1}=\frac{1}{4\pi}\left(1+15 \zeta(3) \lambda^{-3/2}\right)\approx 0.08 + 1.4\lambda^{-3/2}\,,
$$ 
significantly underestimates the slope of the kinetic theory $\eta/s$ values for $\lambda>10$. This behavior has been discussed before in Ref.~\cite{Huot:2006ys}, where it was suggested to change the identification of $\lambda$ when comparing ${\cal N}=4$ SYM to pure Yang-Mills or QCD.

\begin{figure}
\includegraphics[width=0.6\textwidth]{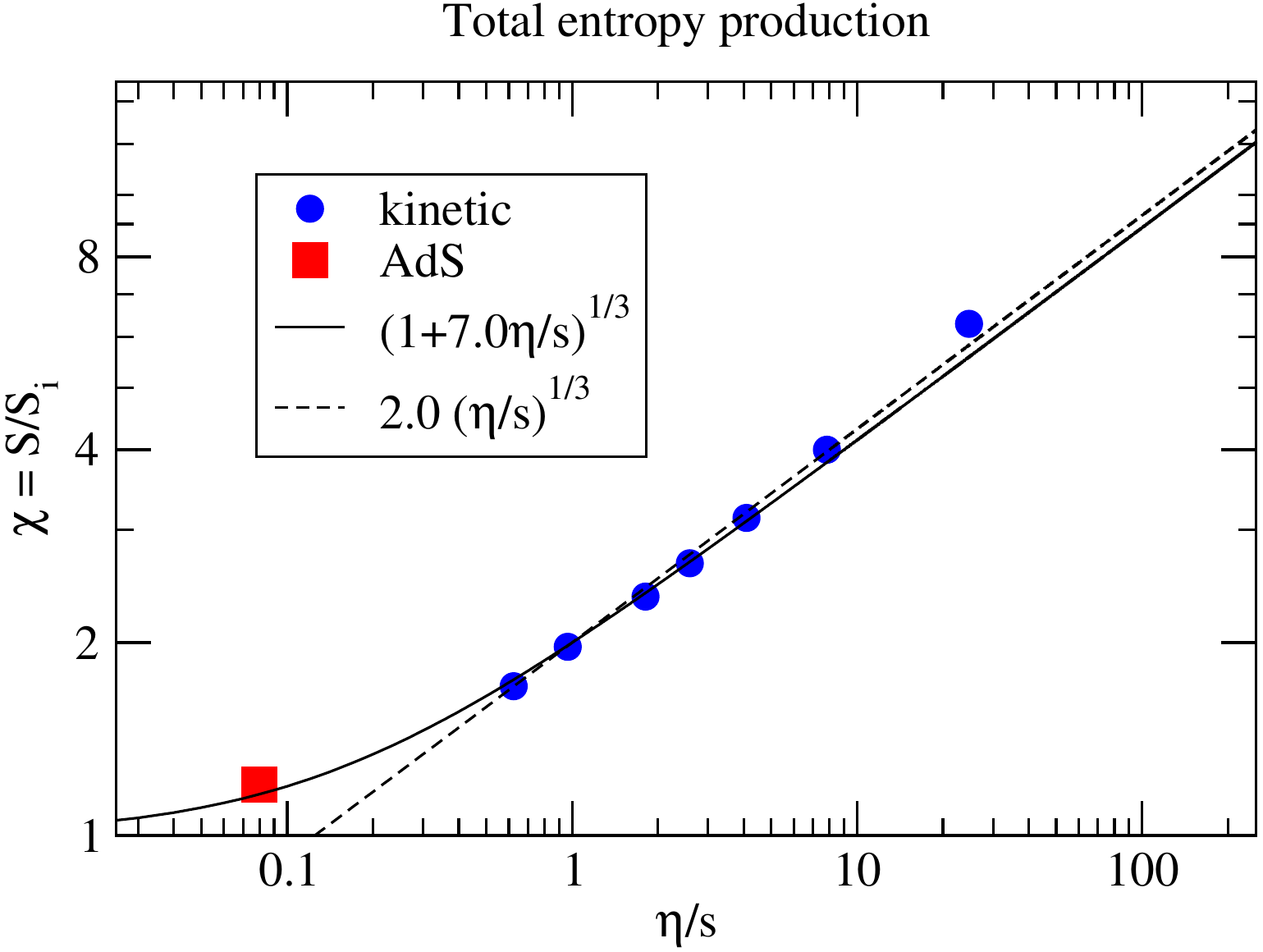}
\caption{\label{plot:chi}
Scaling of the total entropy production during the non-equilibrium evolution. The dashed line corresponds to 
a parametric expectation based on weak coupling picture $\chi\propto (\eta/s)^{1/3}$, while the strong coupling expectation predicts $\chi-1\propto \eta/s$. The empirical result from (\ref{eq:chieff}) (full line) satisfies both limits.  
}
\end{figure}

At weak coupling we may estimate the parametric dependence of the equilibration process as a function $\lambda$ or $\eta/s$. If the coupling is sufficiently small, the system exhibits large scale separations admitting us to parametrically model the evolution as a three stage process:
\begin{itemize}
\item at early times $tT_i< 1$, the system is in thermal equilibrium, 
\item for $1< t T_i <  t_{\rm eq}T_i$ the system exhibits free streaming behaviour and is highly anisotropic $P_L \ll P_\perp$, 
\item and for $ t> t_{\rm eq}$ the system has re-equilibrated and follows inviscid hydrodynamics with $P_L=P_T$. 
\end{itemize}
We expect the system to smoothly change its behaviour from free streaming type evolution to hydrodynamical evolution in the time
scale determined by the transport mean free time, $t_{\rm eq} \sim 1/{\lambda^2 T(t_{\rm eq})}$ %
\footnote{For very small values of $\lambda$ a large scale separation develops between $T_i$ and $T(t_{\rm eq})$ and this estimate should be replaced with the LPM suppressed rate $(T_i/T)^{1/2}/\lambda^2T$ leading to slightly different power laws \cite{Kurkela:2011ti,Kurkela:2015qoa}. Here, in the numerical simulations we do not probe small enough values of $\lambda$ for this to be numerically relevant. }
. Naively the parametric dependence
of this equation is $\propto \lambda^{-2}$, however the expansion reduces the local energy density of the system and therefore also the 
target temperature $T(t_{\rm eq})$ to which the system aspires to thermalize. For a freely streaming anisotropic system the
energy density evolves as $\epsilon(t)\sim \epsilon_i /(T_i t)$, and therefore the target temperature at time $t$ is $T(t) \sim T_i^{3/4}t^{-1/4}$. Solving now 
self-consistently the condition that the system time be of the same order of magnitude as the transport mean free time leads to
\begin{align}
T_i t_{\rm eq} \sim \lambda^{-8/3} \sim (\eta/s)^{4/3}, \quad \quad T(t_{\rm eq}) \sim \lambda^{2/3} T_i \sim (\eta/s)^{-1/3} T_i 
\end{align}
and for the total entropy generation during the second stage
\begin{align}
\label{eq:weakchi}
\chi \sim  T_i t_{eq.} \frac{T(t_{\rm eq})^3}{T_i^3} \sim\frac{1}{\lambda^{2/3}} \sim (\eta/s)^{1/3}\,,
\end{align}
where $\eta/s \propto \lambda^{-2}$ was used.

At strong coupling one has $\frac{\eta}{s}\ll 1$, and as a consequence the viscous entropy production can be calculated from the full hydrodynamic evolution equations (\ref{eq:NSdyn1},\ref{eq:rBRSSSeq}) in an arbitrary background $g(t)$. Specifically, when solving Eq.~(\ref{eq:NSdyn1}) perturbatively in $\frac{\eta}{s}\ll 1$, one finds
\begin{equation}
\label{eq:chistrong}
\chi=1+\frac{\eta}{3 s}\int_{-\infty}^\infty \frac{dt}{T_i}\left(\frac{g^\prime(t)}{g}\right)^{2} g^{1/6}(t)\simeq 1+2.0 \frac{\eta}{s}\,,
\end{equation}
where the specific form of $g(t)$ from Eq.~(\ref{eq:metricfunc}) with $\alpha=8$ was used to calculate the numerical value of $2.0$ in Eq.~(\ref{eq:chistrong}).

Based on the weak and strong coupling results in Eq.~(\ref{eq:weakchi},\ref{eq:chistrong}) for $\chi$, a model function that obeys both these limits is given by
\begin{equation}
\label{eq:chieff}
\chi\simeq \left(1+7.0 \frac{\eta}{s}\right)^{1/3}\,,
\end{equation}
where the value $7.0$ was adjusted to match the results for $\chi$ at weak coupling.

\begin{figure}
\includegraphics[width=0.6\textwidth]{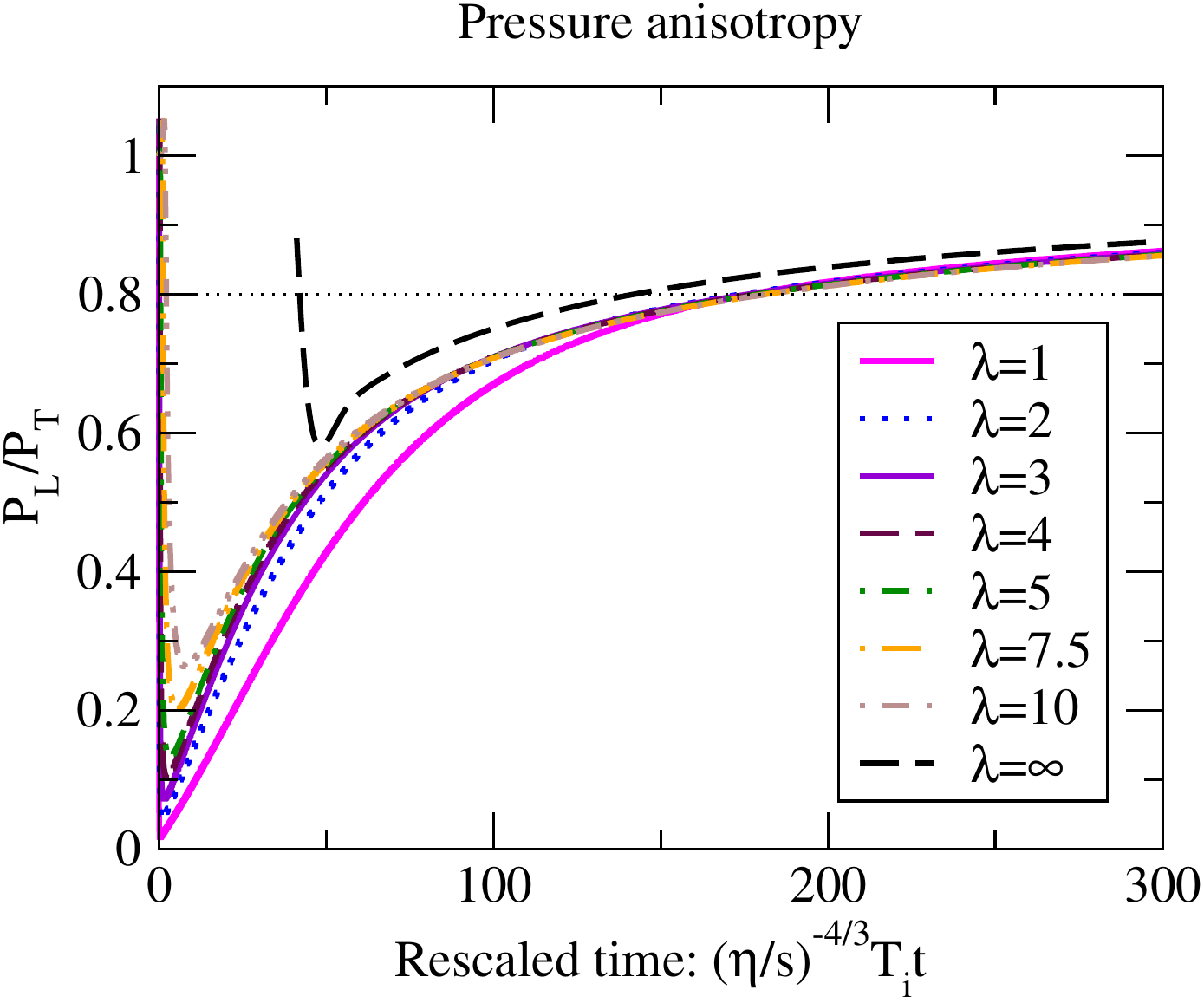}
\caption{
\label{plot:fig5b}
Pressure anistropy with time rescaled by the weak coupling estimate for the thermalization time. All the simulations, including $\lambda=\infty$, follow
a universal attractor, given by Eqn. (\ref{eq:BRSSSplpt2}) , towards thermal equilibrium.
}
\end{figure}

\begin{figure}
\includegraphics[width=0.6\textwidth]{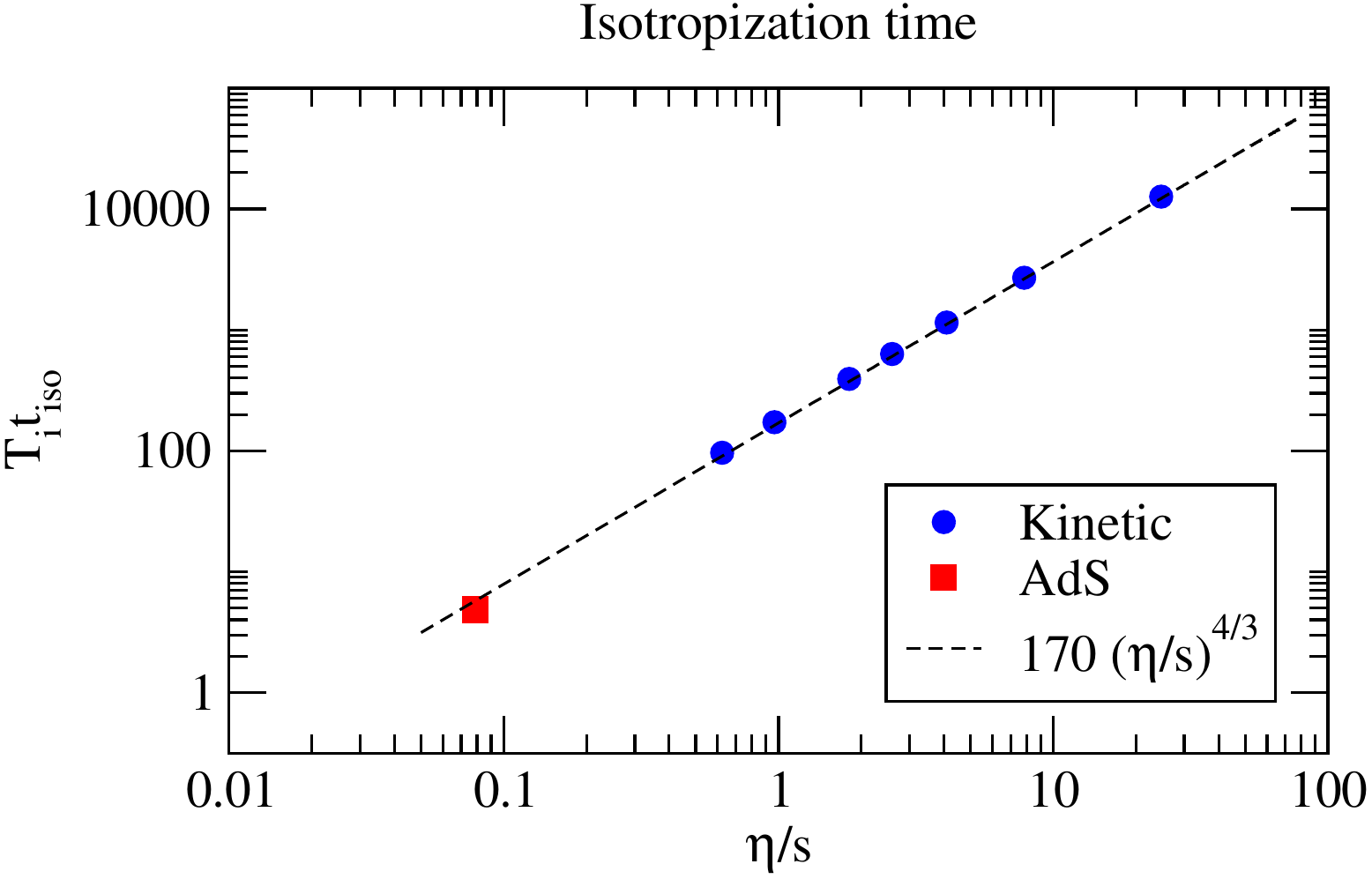}
\caption{
\label{plot:figtiso}
Isotropization time defined by the condition $P_L/P_T = 0.8$ as a function of $\eta/s$.  The parametric model $t_{\rm iso} T_i \propto (\eta/s)^{4/3}$ describes the kinetic theory values extremely well and extrapolates to the strong coupling value within $25\%$.
}
\end{figure}

In Fig.~\ref{plot:chi} we compare the total entropy generation $\chi$ for the weak and strong coupling simulations. We first note
that all points follow a monotonous growing curve.  The parametric model with $(\eta/s)^{1/3}$ describes well the scaling of all  the kinetic theory points. Extrapolating the model to smaller values eventually predicts isentropic 
evolution for $\eta/s \approx 0.13$ when the duration of the second stage goes to zero, thereby clearly signalling the breakdown of the weak coupling picture. It is quite intriguing that the
value where the weak coupling theory predicts its own failure happens to be surprisingly close to the 
strong coupling value of $\eta/s= 0.08$. Unlike the parametric model, strong coupling saturates the interactions and the entropy generation remains finite and positive, as born out by the interpolation function (\ref{eq:chieff}).

Next, we study the isotropization times by plotting the anisotropy ratio $P_L/P_T$ in Fig.~\ref{plot:fig5b}
as function of the rescaled time variable $(\eta/s)^{-4/3} T_i t$. Upon rescaling, all the kinetic theory simulations approximately collapse onto a single curve, and thus all approach
isotropy at same rescaled time. This approach to isotropy can be seen to be governed by viscous hydrodynamics, whereby from (\ref{eq:BRSSSplpt}) it is clear that
\begin{equation}
\label{eq:BRSSSplpt2}
\left.\frac{P_L}{P_T}\right|_{NS,t T_i\gg 1}=1-\frac{8 \eta}{s}\frac{1}{(t T_i)^{2/3} \chi^{1/3}}\approx 1-8\frac{1}{((\eta/s)^{-4/3}t T_i)^{2/3} }\left(\frac{\eta/s }{1+7 \eta/s}\right)^{1/9}\,.
\end{equation}
where we used Eqn. (\ref{eq:chieff}). For large viscosity this formula simplifies, which explains why the weak coupling evolutions follow the universal
attractor shown in Fig.~\ref{plot:fig5b}. Eqn. (\ref{eq:BRSSSplpt2}) can be solved for our fiducial value of 0.8 to give
\begin{equation}
T_i t_{\rm iso} \approx 154\ldots 183\, (\eta/s)^{4/3},
\end{equation}
for a viscosity between $1/4\pi$ and $\infty$.

\begin{figure}
\includegraphics[width=0.5\linewidth]{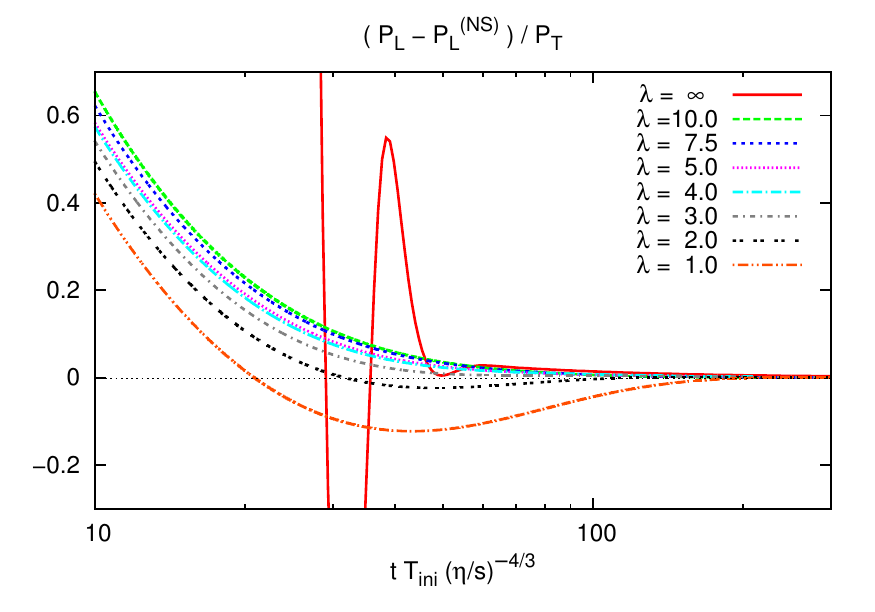}\hfill
\includegraphics[width=0.5\linewidth]{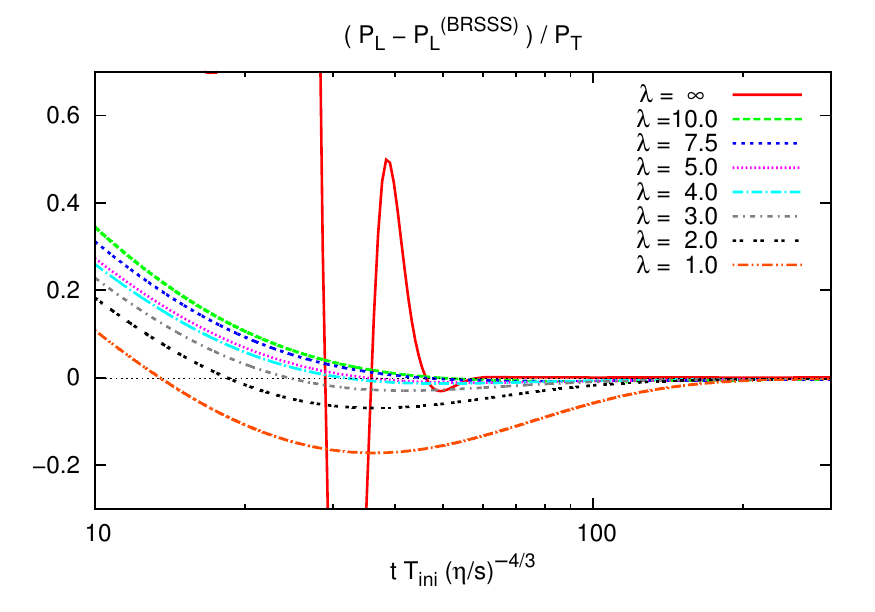}
\caption{
\label{fig5} The deviation of longitudinal pressure from the late time hydrodynamics prediction including the first (left) or the second (right) order terms in the hydrodynamical expansion. 
}
\end{figure}

\begin{figure}
\includegraphics[width=0.6\textwidth]{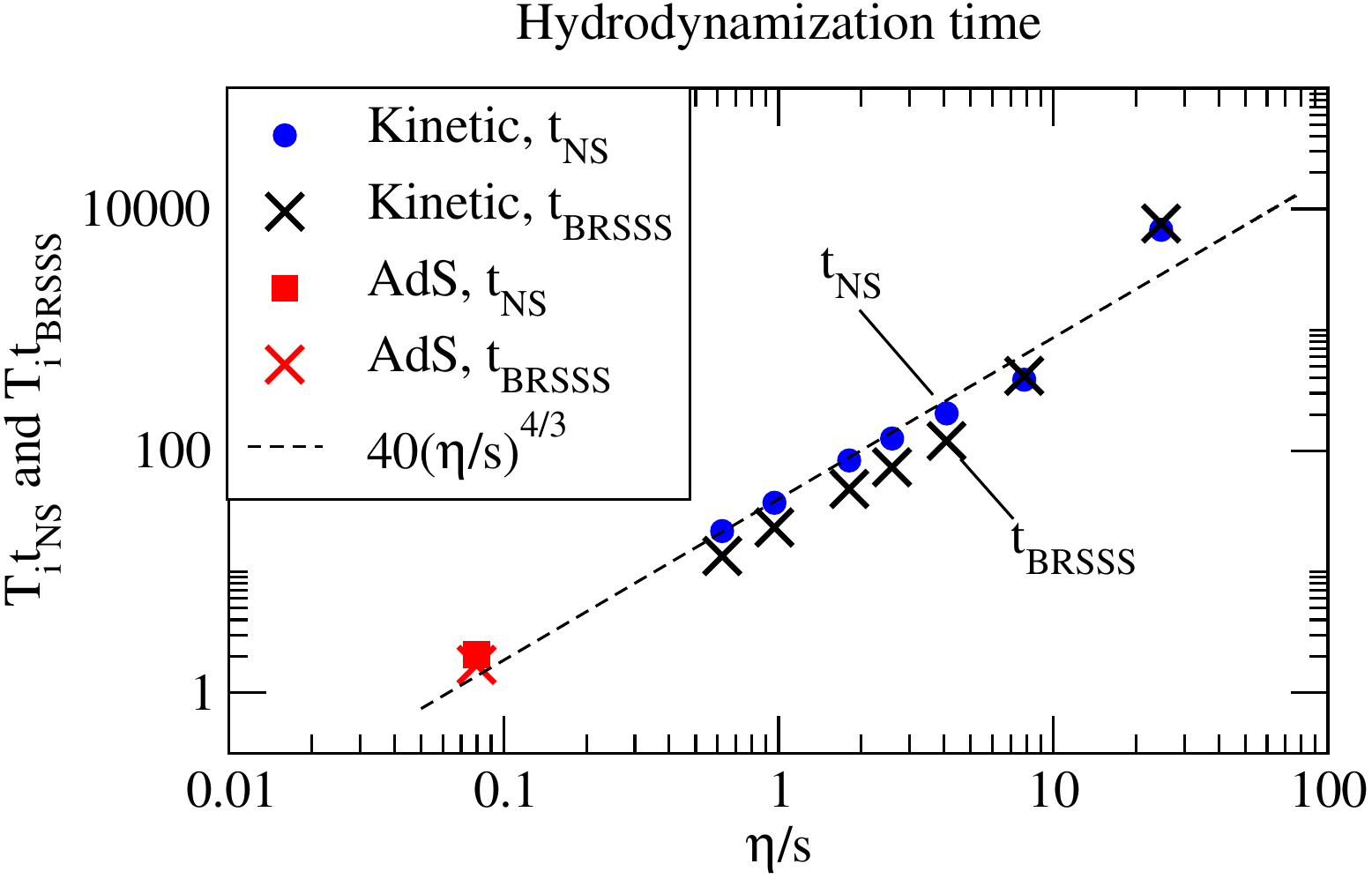}
\caption{
\label{fig:t_NS} Hydrodynamization times as defined by the time when the late time hydrodynamics of Eqs.~(\ref{eq:NStot},\ref{eq:BRSSSplpt}) reproduce the
$P_L/P_T$ ratio of the simulation within $5\%$. The circles correspond to the Navier-Stokes whereas the crosses are second order BRSSS.
}
\end{figure}

Lastly, we focus on hydrodynamization times in Figs. \ref{fig5} and \ref{fig:t_NS}. Fig. \ref{fig5} shows the deviation of $P_L$ from the late time hydro prediction of Eqns.~(\ref{eq:NStot}, \ref{eq:BRSSSplpt}) normalized by the transverse pressure. On the one hand, we again observe that the strong coupling simulation is well described by hydrodynamics immediately after the metric pulse has passed. On the other hand we see that the kinetic theory simulations exhibit a breakdown from hydrodynamics roughly at the same time scale of $T_i t \sim 40 (\eta/s)^{4/3}$. We note that the correspondence with hydrodynamics is slightly improved when the second order coefficients are taken into account, in particular at larger couplings. We note that the exact values of the hydrodynamization times can depend quite strongly on the fiducial range due to the non-monotonic approach to hydrodynamics. Nevertheless, the overall scaling $t_{\rm hyd}\propto (\eta/s)^{4/3}$ remains present even when varying the range. 

We have not yet commented on the generality of our results for different values of $\alpha$ in our metric pulse. We verified that for $\alpha=4$ our results change by less than $1\%$. For significantly faster pulses with $\alpha\gg 8$, however, the calculation starts to differ because of the Hawking radiation generated by the pulse. For $\alpha=16$ this leads for instance to $\chi=1.28$ as compared to $\chi=1.20$ for $\alpha=8$ for $\lambda=\infty$. For the weak coupling framework we did not include Hawking radiation, which is indeed not needed for the profiles we considered.

\section{Conclusions}
\label{sec:conclusions}

In the present work we have presented a detailed and consistent comparison of the equilibration process of a gauge theory at strong and weak coupling using the same set-up and analysis procedure. Our main conclusion is that while there certainly are differences in the thermalization of these two very different theories, there are also some surprising similarities.

We find that the weak coupling thermalization process can be characterized with a simple parametric picture predicting the dependence of 
of thermalization time and entropy production as a function of the coupling constant $\lambda$, or equivalently $\eta/s \propto \lambda^{-2}$. Furthermore, 
extrapolating the powerlaw model to strong couplings where the parametric picture fails, it is surprising that we still found qualitative agreement even to strong coupling simulations. While at the quantitative level this may be a numerical coincidence, it demonstrates the overall similarities of the thermalization processes both at weak and at strong coupling.

The present study is probably too simplistic to be directly applicably for heavy-ion phenomenology. However, it provides evidence that treating weak and strong coupling equilibration on the same footing can lead to simple power-law results that smoothly interpolate between weak and strong coupling. Such interpolation functions may be used to effectively estimate viscosity, equilibration time and viscous entropy production (among others) at intermediate values of the coupling where neither the kinetic nor the gauge/gravity approach are applicable. For instance, for our gauge theory with $\lambda\simeq 20$, our present study would predict $\eta/s\simeq 0.3$, a hydrodynamization time of $\tau T_i\simeq 7$ and a viscous entropy production $\chi-1$ of approximately 40 percent. By repeating our methodological approach for a setup applicable to heavy-ion collisions, our goal for future work is to obtain similar quantitative predictions that would then be directly testable when confronted with precision experimental data.

\begin{acknowledgments}
 
We would like to thank H.~Bantilan, P.~Chesler, K. Kajantie, A.~Mazeliauskas, K.~Rajagopal, K.~Skenderis, I.~Takaaki, D.~Teaney, U. Wiedemann, and C.~Wu for fruitful discussions.
This work was supported by the Department of Energy, DOE award No. DE-SC0008132.  WS is supported by the U.S. Department of Energy under grant Contract Number DE-SC0011090.

\end{acknowledgments}

\bibliographystyle{apsrev} 
\bibliography{weakstrongDec16}

\end{document}